\documentclass[preprint]{aastex_mst}
\usepackage{natbib}
\usepackage{mathrsfs}
\usepackage{rotating}
\usepackage{graphicx}
\usepackage{epstopdf}
\usepackage{longtable,lscape}

\newcommand{\etal}{et~al.}

\newcommand{\fn}[1]{\footnote{\scriptsize{#1}}}
\newcommand{\Eqn}[1]{Eq{#1}.}  
\newcommand{\Fig}[1]{Fig{#1}.}  
\newcommand{\Cassit}{\textit{Cassini}}  

\makeatletter
\renewcommand{\LT@makecaption}[3]{%
  \LT@mcol\LT@cols c{\hbox to\z@{\hss\parbox[t]\LTcapwidth{%
    \sbox\@tempboxa{{#1{\normalsize #2 }}#3}%
    \ifdim\wd\@tempboxa>\hsize
      {#1{\normalsize #2 }}#3%
    \else
      \hbox to\hsize{\hfil\box\@tempboxa\hfil}%
    \fi
    \endgraf\vskip\baselineskip}%
  \hss}}}
\makeatother 
\slugcomment{Accepted at AJ}

\shorttitle{Propeller population in Saturn's A Ring}
\shortauthors{Tiscareno \etal}

\begin{document}

\title{The population of propellers in Saturn's A Ring}
\author{Matthew~S.~Tiscareno$^1$\footnote{Corresponding author:  \tt{matthewt@astro.cornell.edu}}, Joseph~A.~Burns$^{1,2}$, Matthew~M.~Hedman$^1$, Carolyn~C.~Porco$^3$}
\affil{$^1$ Department of Astronomy, Cornell University, Ithaca, NY 14853\\$^2$ Department of Theoretical and Applied Mechanics, Cornell University, Ithaca, NY 14853\\$^3$ CICLOPS, Space Science Institute, 4750 Walnut Street, Boulder, CO 80301}

\begin{abstract}
We present an extensive data set of $\sim 150$ localized features from \Cassit{} images of Saturn's Ring~A, a third of which are demonstrated to be persistent by their appearance in multiple images, and half of which are resolved well enough to reveal a characteristic ``propeller'' shape.  We interpret these features as the signatures of small moonlets embedded within the ring, with diameters between 40 and 500~meters.  The lack of significant brightening at high phase angle indicates that they are likely composed primarily of macroscopic particles, rather than dust.  With the exception of two features found exterior to the Encke Gap, these objects are concentrated entirely within three narrow ($\sim 1000$ km) bands in the mid-A Ring that happen to be free from local disturbances from strong density waves.  However, other nearby regions are similarly free of major disturbances but contain no propellers.  It is unclear whether these bands are due to specific events in which a parent body or bodies broke up into the current moonlets, or whether a larger initial moonlet population has been sculpted into bands by other ring processes.
\end{abstract}

\textit{Subject headings:}  planets: rings \\
\indent{}\textit{Running header:}  Propeller population in Saturn's A Ring

\section{Introduction}
Saturn's main rings (particularly Ring~A) were determined by the Voyager Radio Science experiment \citep{Zebker85} to be primarily composed of a distribution of icy particles of diameter $D \gtrsim 1$~cm.  A steep cutoff in the size-distribution was discerned at $D \sim 20$~m, but particles larger than that value could not be probed due to limitations imposed by the radio experiment's carrier wavelength.  At the large end of the particle-size distribution are the two known moonlets embedded in gaps in the outer A~Ring, Pan and Daphnis, of diameters $\sim 28$~km and $\sim 8$~km, respectively \citep{PorcoSci07}.  Nothing was known about the distribution of intermediate-size ring particles, between 20~m and 8~km in diameter, until the first evidence of ``missing-link'' particles was found in very-high-resolution images taken by the \Cassit{} spacecraft during its insertion into Saturn orbit \citep{Propellers06}.  These intermediate-size moonlets are not directly seen, but rather the propeller-shaped disturbances they create in the ring continuum.  This morphology had previously been predicted by numerical simulations \citep{SS00,SSD02,Seiss05}.  The sizes of the perturbing moonlets, subject to some ambiguities in interpretation, were given as $D \sim 100$ meters.  Their surface densities are quite low relative to smaller particles, corroborating the steep cutoff reported by \citet{Zebker85}. 

We here present and analyze a data set of 158 localized features in the A~Ring, many of which are well-enough resolved to reveal the characteristic propeller shape.  Four of these objects are those reported by \citet{Propellers06}, and another eight were first noted by \citet{Sremcevic07}.  Recently, \citet{Espo07} have found evidence for similarly-sized moonlets in the narrow and highly disturbed F~Ring; however, this is not directly applicable to our analysis because there is no particular reason to expect the particle-size distribution of the F~Ring to be simply related to that of the A~Ring. 

Section~\ref{Propellers} gives further background on the nature and interpretation of propellers.  Section~\ref{Observations} summarizes the imaging sequences we used in compiling our data set, and Section~\ref{Analysis} describes the process by which features were identified and characterized with one of two models (``resolved'' or ``unresolved'').  Our results are summarized in Section~\ref{Results}.  Section~\ref{Interpretation} contains further discussion of the interpretation of propeller features.   A full and unabridged presentation of our data set is given as an Appendix.  

\begin{figure}[!t]
\begin{center}
\includegraphics[width=10cm,keepaspectratio=true]{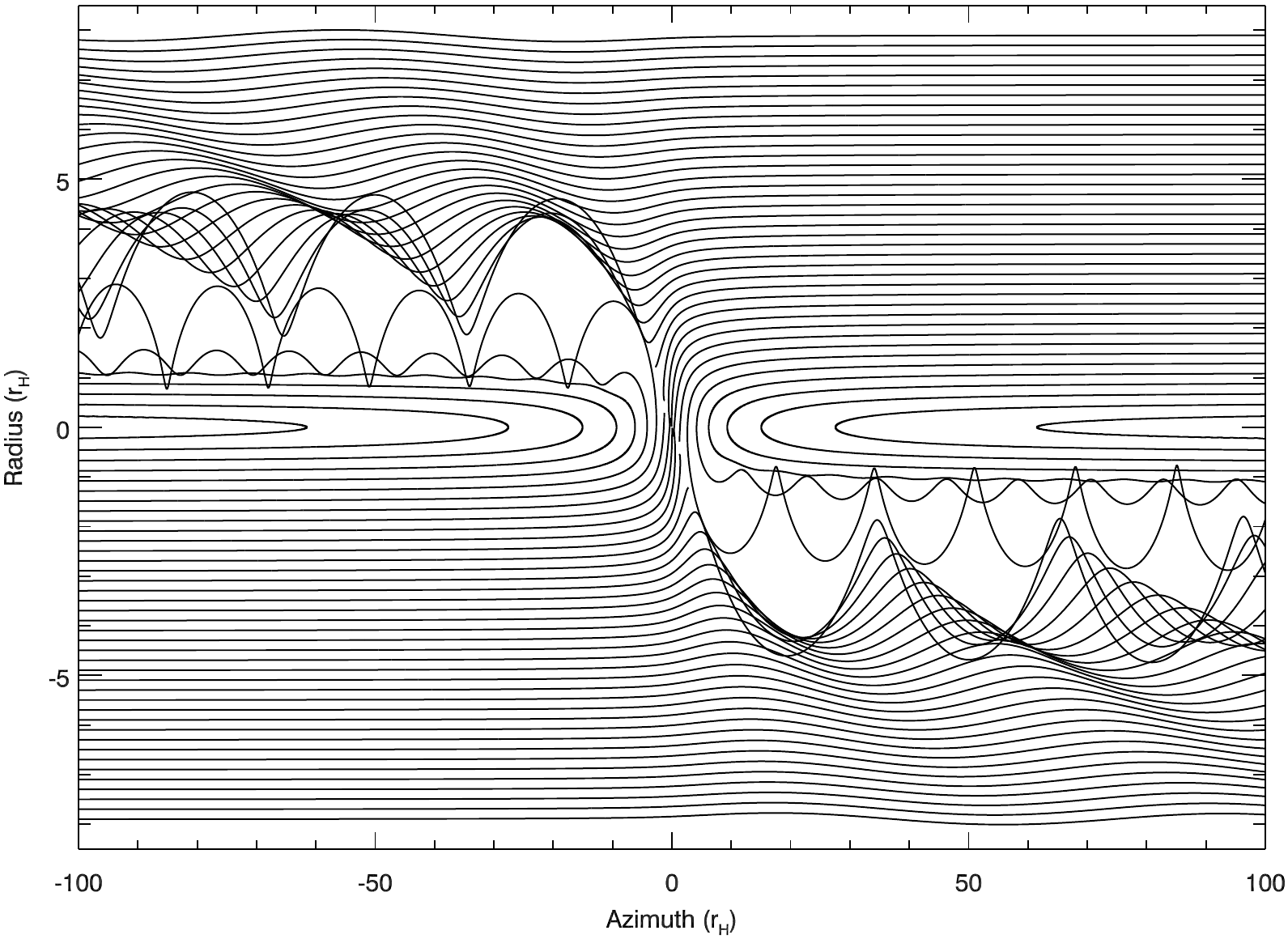}
\caption{Particle trajectories under Hill's equations \citep[see, e.g.,][ch.~3.13]{MD99}, showing the propeller-shaped chaotic zones as well as associated moonlet wakes.  Radial ($r$)) and azimuthal ($\ell$) coordinates are each shown in units of Hill radii; the perturbing mass is located at the origin, and the $L_1$ and $L_2$ Lagrange points are at $\ell = 0$, $r = \pm 1$  The direction towards Saturn is down, and the orbital direction is to the right.  Scattered trajectories deflected by $>5R_H$ are not shown.  This simple model, which neglects inter-particle collisions as well as self-gravity, is presented for conceptual purposes only. \label{Illustration}}
\end{center}
\end{figure}

\section{Propeller-shaped Features \label{Propellers}}
The general morphology of propeller-shaped features, hereafter ``propellers'', can be understood from the classical Hill problem, in which massless test particles orbiting a large central body are deflected in the vicinity of a perturbing mass, hereafter the ``moonlet'' (see \Fig{}~\ref{Illustration}).  It should be noted that this is a simple model for conceptual purposes only, as perturbing effects such as inter-particle collisions and self-gravity wakes (see below) complicate the actual picture considerably.  

Due to keplerian shear, the chaotic zone in which test particles are cleared out by the moonlet is carried forward on the inward side, while that on the outward side trails behind.  The clearing of particles from these chaotic zones results in two gaps, oriented primarily along the orbital direction, symmetric about the origin in the shape of a propeller (\Fig{}~\ref{Illustration}).  The separation between these gaps in the radial direction is approximately 4 times the Hill radius\fn{The Hill radius, a measure of the sphere of a body's gravitational dominance, is given by $r_H = a (m/3M_S)^{1/3}$, where $m$ and $a$ are the moonlet's mass and distance from Saturn, and $M_S$ is Saturn's mass.}.  Herafter, we will refer to this aspect of a moonlet's disturbance, which is a local decrease in the total surface density, as \textit{\textbf{propeller-shaped gaps}}.  

Regions of enhanced surface density can also be created by the perturbing moonlet.  The genesis of these density enhancements, as is easily seen in the non-self-gravitating case shown in \Fig{}~\ref{Illustration}, is essentially the same as the ``moonlet wakes'' that are seen elsewhere in the rings on a larger scale --- for example, in the vicinity of Pan \citep{Show86}.  Hereafter, we will refer to this aspect of a moonlet's disturbance as \textit{\textbf{moonlet wakes}}.  These are farther from the moonlet, in terms of radial displacement, than are the gaps.  They also differ from the gaps in their general morphology, being less consistently oriented in the azimuthal direction, though this distinction could possibly be less important in more realistic cases that account for self-gravity wakes.  

In reality, the motion of particles in Saturn's rings is far from the laminar flow portrayed in the Hill problem.  Inter-particle collisions play a significant role, especially at locations where streamlines cross, but even more important are ``self-gravity wakes''\fn{It is unfortunate that the word ``wake'' is applied to two very different ring processes, moonlet wakes and self-gravity wakes, both of which are important to the phenomena discussed in this paper.} \citep{JT66,Salo95}, a constant clumping together and shearing apart of ring particles as their mutual gravity is balanced against Saturn's tides.  Self-gravity wakes pervade the A~Ring and significantly affect its photometry \citep{Colwell07,Hedman07,PorcoPhot07}.  In numerical simulations \citep{LS07,LewisDDA07,Sremcevic07}, the wakes produced by considering particle self-gravity cause a propeller's morphology to be altered significantly.  As one changes the particle-size regime as well as the perturbing moonlet size, the strength (i.e., deviation from the continuum background) and the azimuthal length of the propeller feature can vary, and the region of moonlet wakes can change in shape or even disappear.  However, the existence of a propeller-shaped gap and the radial separation between the two lobes have so far appeared in every model in which the perturbing moonlet is large enough to produce any signature at all.  

The morphology of observed propellers fits best with the gaps (regions of diminished surface density) seen in models, with two unitary offset lobes that are oriented precisely in the azimuthal direction \citep{Propellers06}.  However, the fact that propellers are observed as bright features relative to background requires us to consider the possibility that the observed features are caused by moonlet wakes, rather than gaps.  We will refer to this ambiguity in Section~\ref{Results}, and discuss it explicitly in Section~\ref{Interpretation}. 

\section{Observations \label{Observations}}
The images in which our propellers have been found were taken by the Imaging Science Subsystem \citep{PorcoSSR04} on board the \Cassit{} spacecraft, on several occasions during its tour of the Saturn system.  The viewing geometry is described in Table~\ref{ObserveInfo}, in which images taken during the same observing sequence are grouped together.  The two discovery images \citep{Propellers06}, taken during \Cassit{}'s Saturn Orbit Insertion maneuver (SOI) upon its arrival at Saturn on 1~July~2004 (day-of-year 183), remain the highest-resolution images ever obtained of the rings.  Propellers were next found in a suite of images taken in Orbit~13 during an occultation by the rings of the star $\alpha$~Scorpii, an event planned by \Cassit{}'s UVIS and VIMS spectrometers \citep{Hedman05,Colwell06,Colwell07,Espo07,Sremcevic07}.  Also in Orbit~13, a propeller was captured in one of a series of 105 images of the Encke Gap.  In each of these cases, the propellers were seen in diffuse transmission, with the camera on the unlit side of the rings.  More recently, propellers have been captured during four complete, high-resolution radial scans of the sunlit side of the rings at high resolution during Orbits~28, 31, 32 and~46.  

Other images were taken during the SOI maneuver in which propellers might have been detected but were not.  A second pair of ultra-high-resolution ($\sim 0.05$~km/pixel) images, similar to the discovery images, was taken of the mid-A~Ring with similar emission and phase angles ($16.5^\circ$ and $130.6^\circ$, respectively).  These are probably significant non-detections; the footprint of one (N1467347504) lies at 132,580~km, just beyond the outer edge of the propeller belt (see Section~\ref{Locs}), while that of the other (N1467347445) falls within the moderate-strength Prometheus 12:11 spiral density wave (see Section~\ref{Locs}).  The primary SOI data set was a series of 39 high-resolution images, incompletely covering the rings at resolutions of 0.16~to 0.28~km/pixel \citep{Porco05,soirings}; these images have somewhat higher emission and lower phase angles than the ultra-high-resolution images, but may lack propellers primarily because of the low signal-to-noise that was necessary to prevent smear due to high spacecraft velocity.  

\begin{center}
\begin{table}[!t]
\begin{scriptsize}
\begin{center}
\caption{Observing information for images used in this paper. \label{ObserveInfo}}
\begin{tabular} { l c c c c c c c c }
\hline
\hline
\vspace{-0.035in}
 & & & & Incidence & Emission & Phase & Radial & Azimuthal \\
Image & Orbit & Num$^a$ & Date/Time (UT) & Angle$^b$ & Angle$^b$ & Angle & Resolution$^c$ & Resolution$^c$ \\
\hline
N1467347210 & SOI & 41 & 2004-183 04:02:35 & 114.5$^\circ$ &   5.3$^\circ$ & 119.0$^\circ$ & 0.05 & 0.05 \\
N1467347249 & SOI & 42 & 2004-183 04:03:14 & 114.5$^\circ$ &   5.4$^\circ$ & 119.0$^\circ$ & 0.05 & 0.05 \\
\hline
N1503229987 & 013 & 08 & 2005-232 11:25:00 & 110.7$^\circ$ &  57.9$^\circ$ & 127.8$^\circ$ & 0.97 & 0.59 \\
N1503230047 & 013 & 09 & 2005-232 11:26:00 & 110.7$^\circ$ &  57.9$^\circ$ & 127.8$^\circ$ & 0.98 & 0.60 \\
N1503230107 & 013 & 10 & 2005-232 11:27:00 & 110.7$^\circ$ &  57.9$^\circ$ & 127.8$^\circ$ & 0.99 & 0.60 \\
N1503230167 & 013 & 11 & 2005-232 11:28:00 & 110.7$^\circ$ &  57.9$^\circ$ & 127.8$^\circ$ & 1.00 & 0.60 \\
N1503230227 & 013 & 12 & 2005-232 11:29:00 & 110.7$^\circ$ &  57.9$^\circ$ & 127.8$^\circ$ & 1.01 & 0.61 \\
N1503230287 & 013 & 13 & 2005-232 11:30:00 & 110.7$^\circ$ &  57.9$^\circ$ & 127.8$^\circ$ & 1.02 & 0.61 \\
N1503230347 & 013 & 14 & 2005-232 11:31:00 & 110.7$^\circ$ &  57.9$^\circ$ & 127.8$^\circ$ & 1.03 & 0.61 \\
N1503230407 & 013 & 15 & 2005-232 11:32:00 & 110.7$^\circ$ &  57.9$^\circ$ & 127.8$^\circ$ & 1.04 & 0.61 \\
N1503230467 & 013 & 16 & 2005-232 11:33:00 & 110.7$^\circ$ &  57.9$^\circ$ & 127.8$^\circ$ & 1.05 & 0.62 \\
\hline
N1503243458 & 013-AZ$^d$ & 20 & 2005-232 15:09:30 & 110.7$^\circ$ &  52.1$^\circ$ & 162.3$^\circ$ & 1.32 & 1.10 \\
\hline
N1536497993 & 028 & 30 & 2006-252 12:28:12 & 105.9$^\circ$ & 120.1$^\circ$ &  49.9$^\circ$ & 1.43 & 2.33 \\
N1536498176 & 028 & 32 & 2006-252 12:31:15 & 105.9$^\circ$ & 120.1$^\circ$ &  49.9$^\circ$ & 1.43 & 2.33 \\
N1536498268 & 028 & 33 & 2006-252 12:32:47 & 105.9$^\circ$ & 120.0$^\circ$ &  49.9$^\circ$ & 1.43 & 2.33 \\
N1536498361 & 028 & 34 & 2006-252 12:34:20 & 105.9$^\circ$ & 120.0$^\circ$ &  49.9$^\circ$ & 1.43 & 2.33 \\
N1536498453 & 028 & 35 & 2006-252 12:35:52 & 105.9$^\circ$ & 120.0$^\circ$ &  49.9$^\circ$ & 1.43 & 2.33 \\
\hline
N1540681073 & 031 & 47 & 2006-300 22:25:46 & 105.3$^\circ$ & 153.9$^\circ$ &  48.7$^\circ$ & 1.36 & 1.22 \\
\hline
N1541716008 & 032 & 44 & 2006-312 21:54:34 & 105.1$^\circ$ & 155.9$^\circ$ &  51.0$^\circ$ & 1.40 & 1.28 \\
N1541716180 & 032 & 45 & 2006-312 21:57:26 & 105.1$^\circ$ & 155.8$^\circ$ &  50.9$^\circ$ & 1.39 & 1.27 \\
N1541716352 & 032 & 46 & 2006-312 22:00:18 & 105.1$^\circ$ & 155.7$^\circ$ &  50.8$^\circ$ & 1.38 & 1.26 \\
\hline
N1560310219 & 046 & 07 & 2007-163 02:56:07 & 102.1$^\circ$ & 129.2$^\circ$ &  46.5$^\circ$ & 0.67 & 0.73 \\
N1560310460 & 046 & 09 & 2007-163 03:00:08 & 102.1$^\circ$ & 128.4$^\circ$ &  46.0$^\circ$ & 0.69 & 0.76 \\
N1560310609 & 046 & 10 & 2007-163 03:02:37 & 102.1$^\circ$ & 127.9$^\circ$ &  45.6$^\circ$ & 0.71 & 0.77 \\
N1560310728 & 046 & 11 & 2007-163 03:04:36 & 102.1$^\circ$ & 127.5$^\circ$ &  45.4$^\circ$ & 0.72 & 0.79 \\
N1560310846 & 046 & 12 & 2007-163 03:06:34 & 102.1$^\circ$ & 127.0$^\circ$ &  45.1$^\circ$ & 0.74 & 0.80 \\
\hline
\end{tabular}
\end{center}
\begin{flushleft}
\vspace{-0.1in}
$^a$ An internal number identifying each image within its observing sequence;  used for naming features in Table~\ref{PropTable}. \\
$^b$ Measured from the direction of Saturn's north pole (ring-plane normal).  Note that images from SOI and Orbit 13 view the unlit face of the rings, the others the lit face.  \\
$^c$ In km/pixel.\\
$^d$ This designation is used to identify a second observing sequence in Orbit~13 that found a propeller; this one is an azimuthal scan rather than a radial one.  See Sections~\ref{Observations} and~\ref{Abundance}. 
\end{flushleft}
\end{scriptsize}
\end{table}
\end{center}

\begin{figure}[!t]
\begin{center}
\includegraphics[width=16cm,keepaspectratio=true]{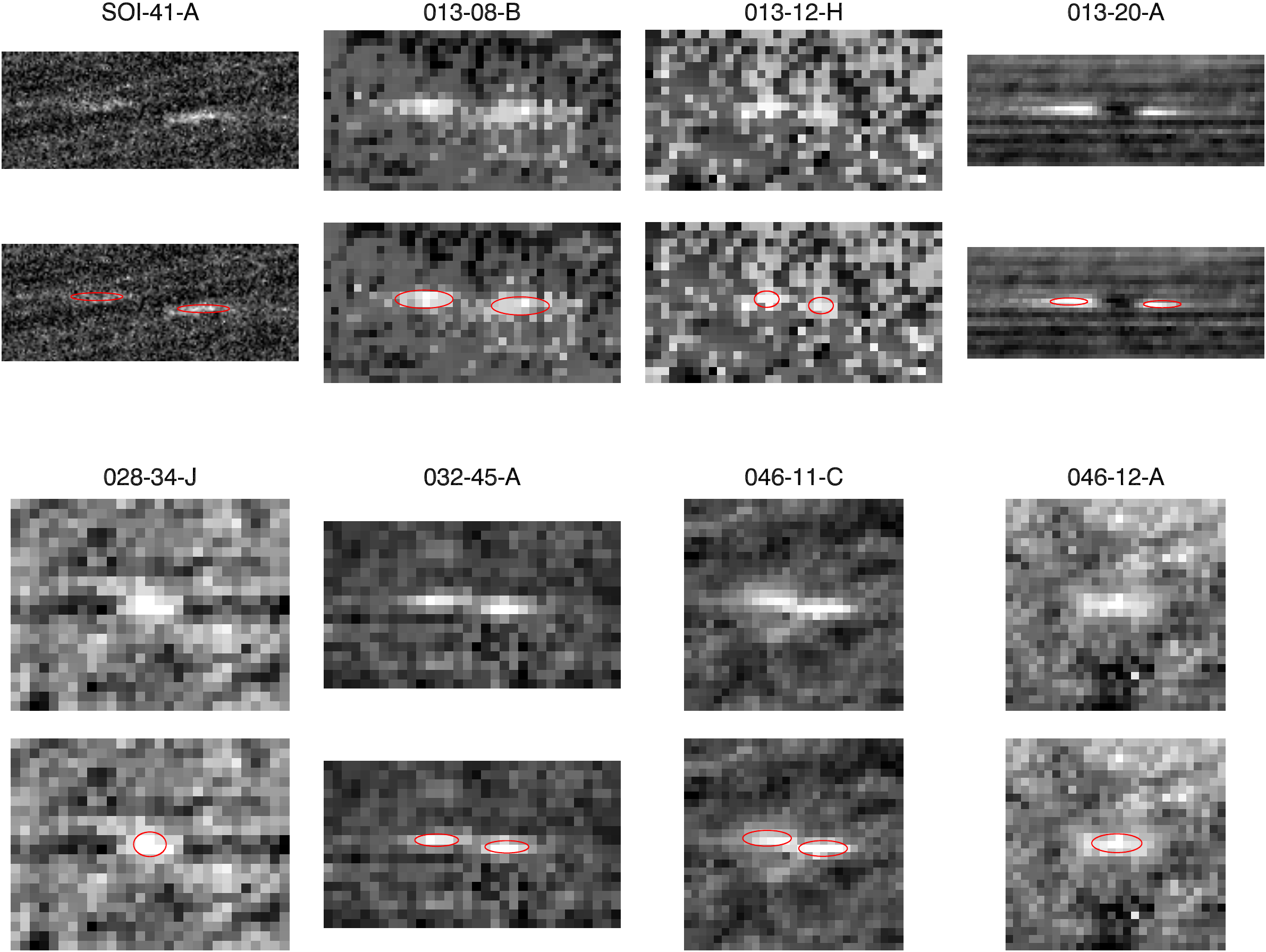}
\caption{Selected examples of features in our data set, shown both with and without the model fit overplotted.  The direction towards Saturn is down, and the orbital direction is to the right.  Six of the features shown here are fit with the primary propeller model (\Eqn{}~\ref{DoubleGaussianModel}), while 028-34-J and 046-12-A are fit with the secondary model (\Eqn{}~\ref{SingleGaussianModel}).  Each of these 8 features is also shown in \Fig{}~\ref{PropReproj}, and fit parameters are listed in Table~\ref{PropTable}.  \label{PropExamples}}
\end{center}
\end{figure}

\section{Data Analysis \label{Analysis}}
\begin{figure}[!t]
\begin{center}
\includegraphics[width=16cm,keepaspectratio=true]{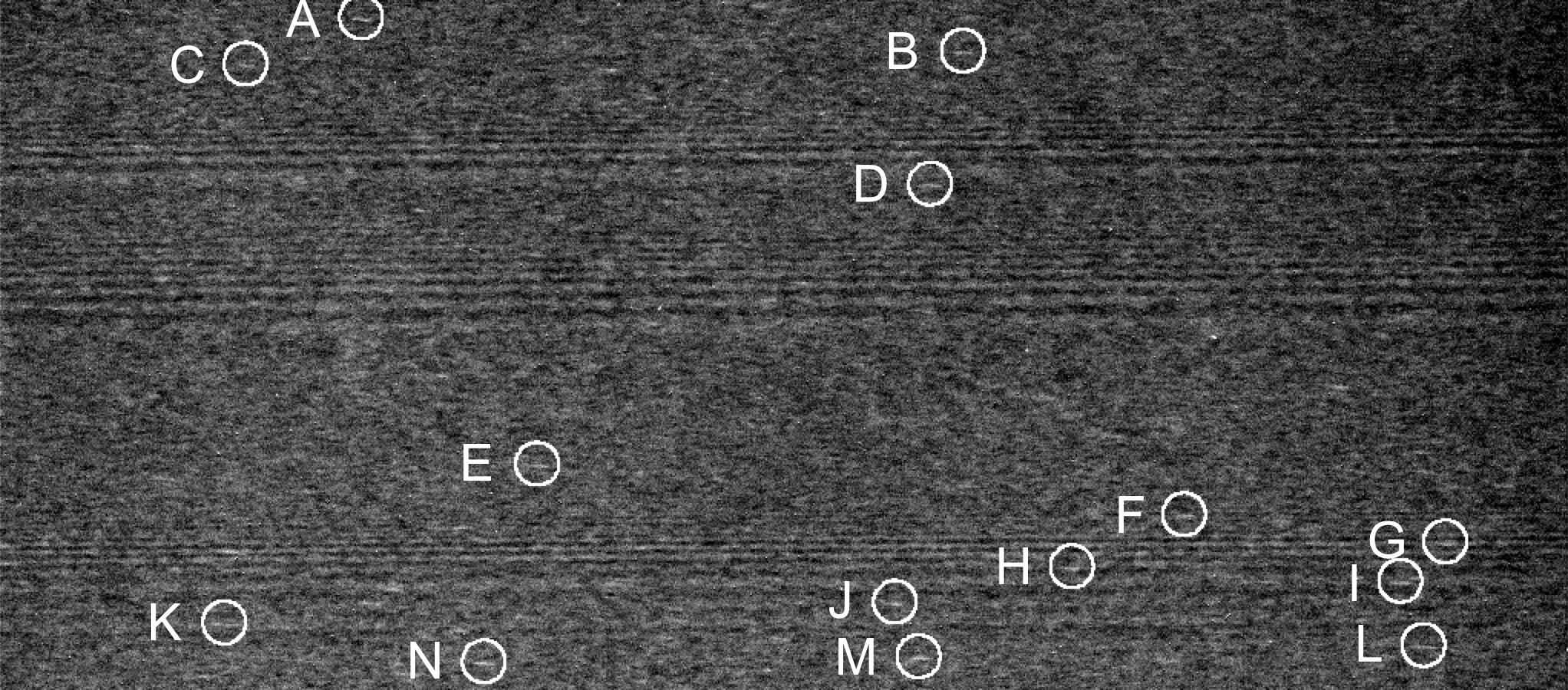}
\includegraphics[width=16cm,keepaspectratio=true,viewport=0.05cm 0cm 25.95cm 9cm]{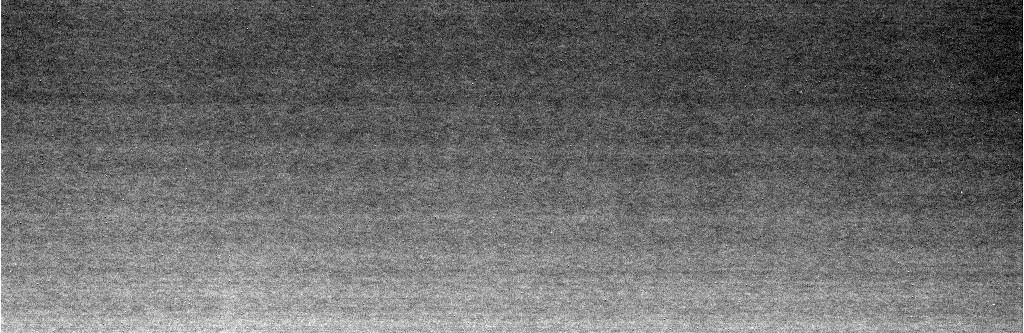}
\caption{(top) Portion of \Cassit{} image N1560310609 (from Orbit~46), showing a region centered at 129,250 km from Saturn's center.  An azimuthally-aligned clumpy background texture is apparent, along with propellers 046-10-A through 046-10-N as identified in Table~\ref{PropTable}.  Also seen are three weak spiral density waves (from bottom to top: Pan 20:19, Atlas 11:10, and Pan 21:20).  (bottom) Portion of \Cassit{} image N1560311199, from the same observation sequence but centered at 125,000~km from Saturn's center.  Neither azimuthally-aligned texture nor propellers are seen.  The regular horizontal banding is a camera artifact \citep{PorcoSSR04}.  \label{CompareImages}}
\end{center}
\end{figure}

The images were calibrated using standard procedures \citep{PorcoSSR04}, resulting in pixel values in units of $I/F$ (reflected or transmitted brightness normalized by the incident solar flux).  Candidate propellers were then identified by eye; the typical morphology is a localized feature that is significantly larger in size than any background azimuthal structure (e.g., see \Fig{}~\ref{CompareImages}) and not obviously an image artifact or cosmic ray.  Identified candidates were all bright relative to image background and were typically oriented in the azimuthal direction, though neither of these was a criterion for initial selection. 

Subject to image-exposure conditions, some candidate propellers are quite faint --- those from SOI and Orbit~13, in particular, are only a few photodetector counts (data numbers) above the background --- requiring the image brightness to be ``stretched'' significantly for them to be clearly visible.  The remnants of imperfect removal of the faint 2-Hz horizontal banding structure characteristic of the ISS camera \citep{PorcoSSR04} is sometimes visible alongside these fainter propellers.  

Each candidate was reprojected onto a radius-azimuth grid (see \Fig{s}~\ref{PropExamples} and~\ref{PropReproj}).  The data were then subjected to a 2-dimensional Levenberg-Marquardt least-squares fit \citep{NumericalRecipes} to one of two models.  

\begin{figure}[!t]
\begin{center}
\includegraphics[width=10cm,keepaspectratio=true]{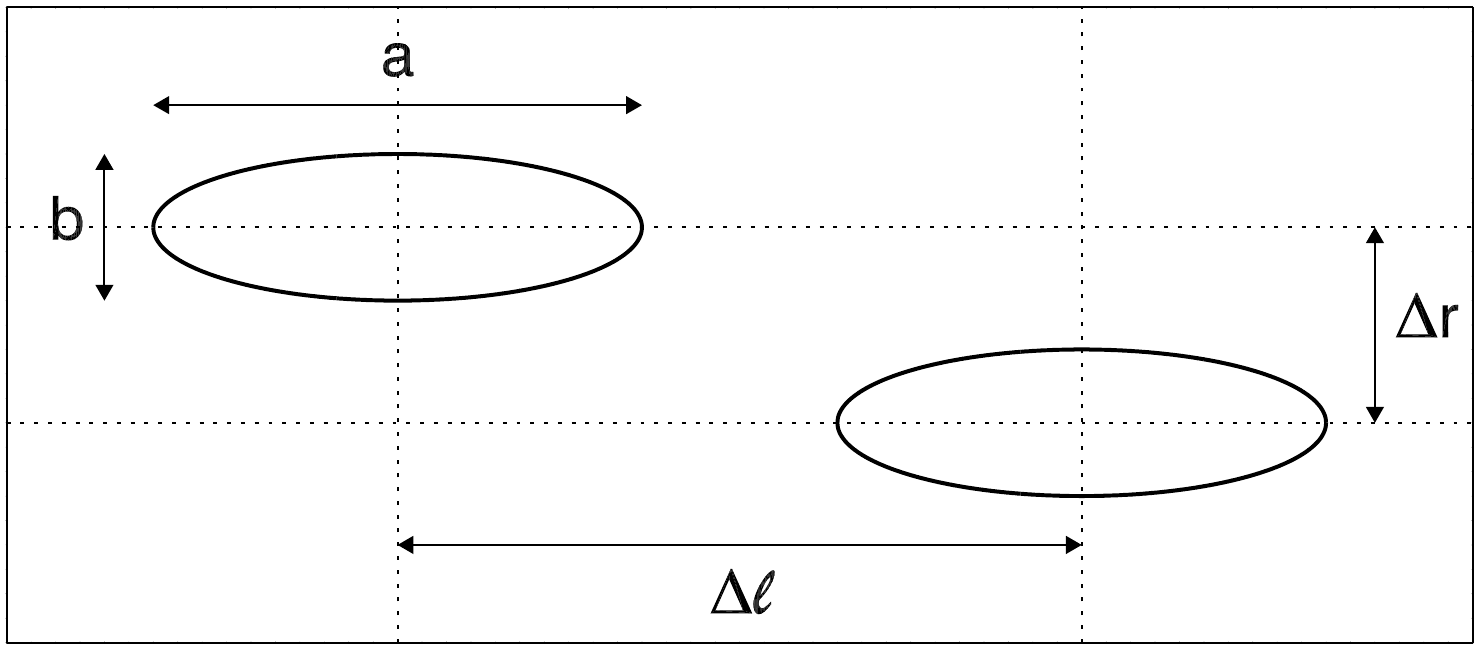}
\caption{The propeller model described by \Eqn{}~\ref{DoubleGaussianModel}.  The coordinate system is a radius-azimuth grid, in which the direction towards saturn is down ($-r$) and the orbital direction is to the right ($+\ell$).  \label{PropModel}}
\end{center}
\end{figure}

The primary model,
{\setlength\arraycolsep{0pt}
\begin{eqnarray}
\label{DoubleGaussianModel}
f(\ell,r) = A_0 + A_1 & \Bigg[ & 
\exp \left\{ -\frac{1}{2} \left( \left[ \frac{\ell-\ell_0+\Delta\ell/2}{a/2} \right]^2 + \left[ \frac{r-r_0-\Delta r/2}{b/2} \right]^2 \right) \right\} + \nonumber \\
&& 
\exp \left\{ -\frac{1}{2} \left( \left[ \frac{\ell-\ell_0-\Delta\ell/2}{a/2} \right]^2 + \left[ \frac{r-r_0+\Delta r/2}{b/2} \right]^2 \right) \right\} \Bigg] , 
\end{eqnarray}}consists of two gaussian peaks of width $a$ in the azimuthal direction and $b$ in the radial direction, with the two identical peaks whose centers are separated by a distance $\Delta \ell$ in the azimuthal direction and $\Delta r$ in the radial direction \citep[see also][]{Sremcevic07}.  A well-resolved propeller should have $\Delta r > 0$ (the leading arm is closer to Saturn than the trailing arm) and $(\Delta \ell - a) > 0$ (the two arms do not overlap azimuthally).  The background brightness is $A_0$, the brightness amplitude is $A_1$, and the center of the model is located at $[\ell_0, r_0]$.  \Fig{}~\ref{PropModel} illustrates this model.  

For cases in which the feature is not well-enough resolved to provide a robust fit to \Eqn{}~\ref{DoubleGaussianModel}, a secondary model consists of a single gaussian peak, of width $a$ in the azimuthal direction and $b$ in the radial direction, centered at $[\ell_0, r_0]$: 
\begin{equation}
\label{SingleGaussianModel}
f(\ell,r) = A_0 + A_1 \exp \left\{ -\frac{1}{2} \left( \left[ \frac{\ell-\ell_0}{a/2} \right]^2 + \left[ \frac{r-r_0}{b/2} \right]^2 \right) \right\} . 
\end{equation}

\Fig{}~\ref{PropExamples} shows a representative sample of the features in our data set and the results of the model fits; the full data set is shown in \Fig{}~\ref{PropReproj} in the Appendix. 

\begin{figure}[!t]
\begin{center}
\includegraphics[width=8cm,keepaspectratio=true]{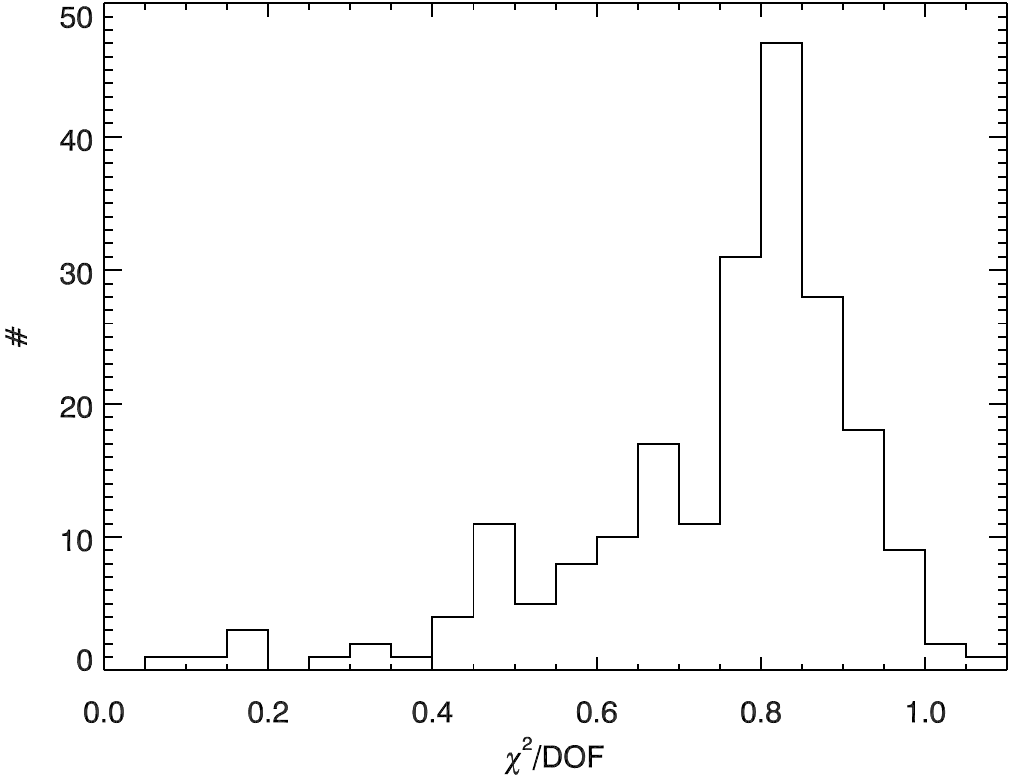}
\caption{Goodness-of-fit statistics for model fits given in Table~\ref{PropTable}.  \label{ChisqPlot}}
\end{center}
\end{figure}

The Levenberg-Marquardt algorithm yields an eight-parameter fit for the primary model (or a six-parameter fit for the secondary model), with error estimates for each parameter and a $\chi^2$ value characterizing the goodness of the fit.  A histogram of reduced $\chi^2$ values is shown in \Fig{}~\ref{ChisqPlot}; they are consistently less than unity, with a median value of 0.78.  Fitted parameters are listed in Table~\ref{PropTable} in the Appendix. 

In order to keep track of the larger number of features in our data set, we use a more specific nomenclature system than previous work (see Table~\ref{PropTable} for details).  For propellers that appear in multiple images, each apparition is given its own name, though the correlation is noted in \Fig{}~\ref{PropReproj} and Table~\ref{PropTable}.  Features noted in previous work are related to our nomenclature in Table~\ref{SremNames}.  

\begin{center}
\begin{table}[!t]
\begin{scriptsize}
\begin{center}
\caption{Correspondence among feature designations in this paper and previous work.  \label{SremNames}}
\begin{tabular} { c c c }
\hline
\hline
\citet{Propellers06} & \citet{Sremcevic07} & This Work \\
\hline
Feature 1 & SOI1 & SOI-41-A \\
Feature 2 & SOI2 & SOI-42-A \\
Feature 3 & SOI3 & SOI-42-B \\
Feature 4 & SOI4 & SOI-42-C \\
& A & 013-08-G / 013-09-C \\
& B & 013-08-F / 013-09-B \\
& C & 013-08-D \\
& D & 013-08-B \\
& E & 013-12-O / 013-13-F \\
& F & 013-12-N / 013-13-E \\
& G & 013-12-P / 013-13-G \\
\hline
\end{tabular}
\end{center}
\end{scriptsize}
\end{table}
\end{center}

\section{Results \label{Results}}
Of the initial 267 candidate propellers identified by eye, 111~(42\%) were at a co-orbiting location that also appeared in an adjacent image (for all numbers given in this section, we do \textit{not} double-count features that appear in multiple images).  31~of these matching locations (28\%) were independently identified by eye as candidates before the correlations among images were noted.  Upon completion of the full analysis of all candidates (including  all matching locations in adjacent images, whether or not those matching locations were independently identified as candidates), 158~features are presented in Table~\ref{PropTable}, 51~of which (33\%) appear in multiple images.  The latter occur entirely in the data from Orbit~13 (27) and Orbit~28 (19), which are the only observing sequences included here in which adjacent images overlap in co-orbiting longitude. 

Of the 158 reported features, 78~of them (49\%) can be fit to the primary model (Section~\ref{Analysis}), indicating that they are well-resolved as propellers.  This includes 20 propellers that appear in two adjacent images and are well-resolved in each, and another 7 that are well-resolved in only one of the images in which they appear, for a total of 27 resolved features (52\%) among the 52 appearing in multiple images.  The ability to resolve the detailed shape of an object is, of course, highly dependent on image resolution and other observational parameters (see Table~\ref{ObserveInfo}) as well as the object's size.  The high-resolution data from Orbit~13 and Orbit~46 allow 84\% and 64\% (respectively) of features to be fit with the primary model, while only 2\% of features in the lower-quality Orbit~28 data are resolved. 

\begin{figure}[!p]
\begin{center}
\includegraphics[width=14cm,keepaspectratio=true]{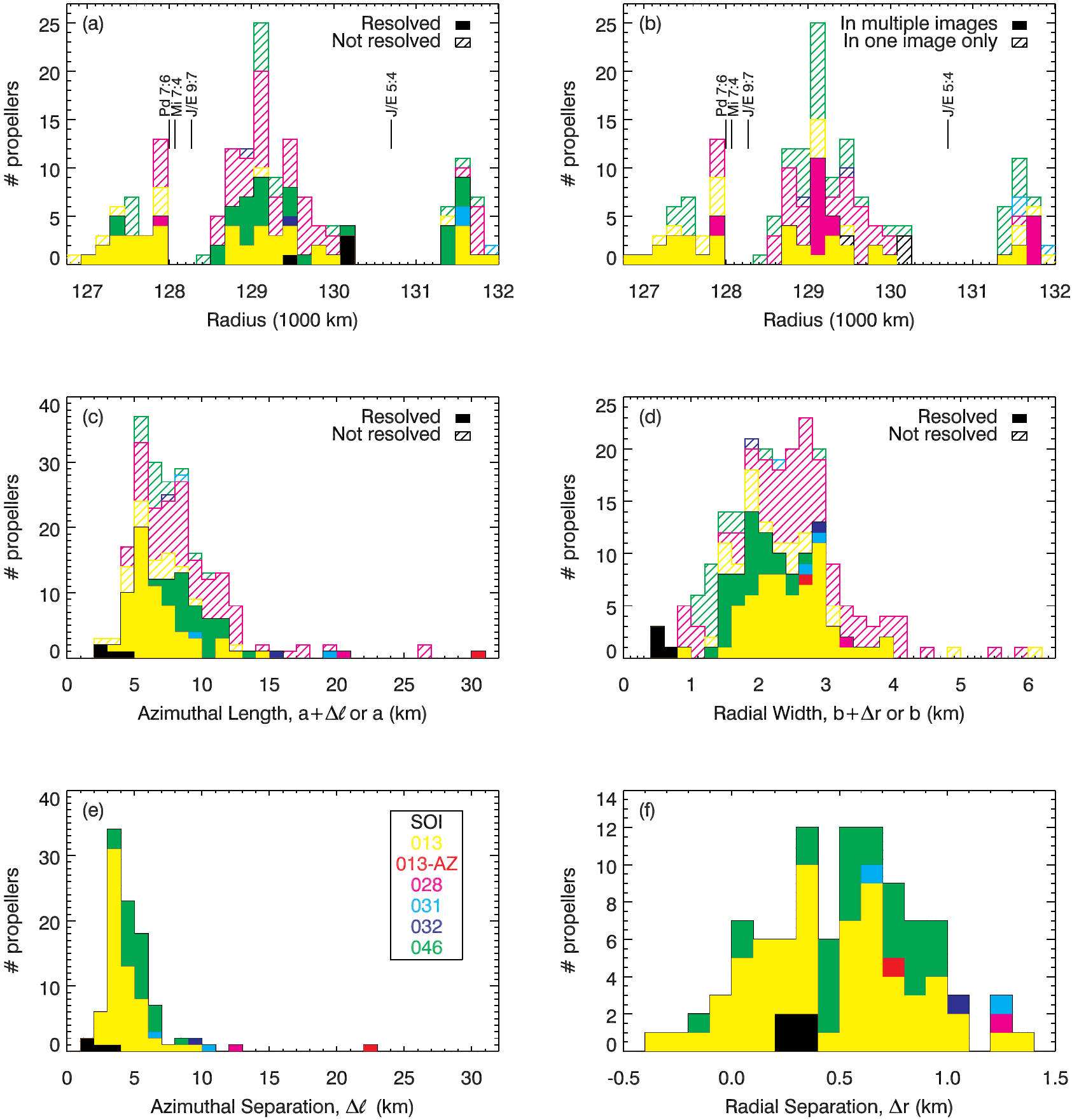}
\caption{In every panel, colors indicate image sequence (see legend in panel ($e$) and Table~\ref{ObserveInfo}).  ($a$) The number of features reported in this paper plotted versus location in the rings (distance from Saturn's center); solid bars indicate resolved propellers, cross-hatched bars unresolved features.  Locations of density waves discussed in Section~\ref{Locs} are indicated.  ($b$) Same as above, except solid bars now indicate features appearing in multiple images.  Off the edge of the plot and not shown in these two panels are features 013-20-A (134,082~km) and 028-30-A (134,715~km).  ($c$) Total azimuthal length, $a+\Delta \ell$ for resolved features, $a$ for unresolved features.  ($d$) Total radial width, $b+\Delta r$ for resolved features, $b$ for unresolved features.  The quantities in these two panels are useful for comparing resolved and unresolved features on the same scale.  ($e$) Azimuthal separation between lobe centers for resolved propellers, $\Delta \ell$.  ($f$) Radial separation between lobe centers for resolved propellers, $\Delta r$; in this panel the scatter (including a few features with $\Delta r < 0$) is dominated by measurement error.  The quantities in the latter two panels are ideally independent of image resolution.  \label{prop_locs}}
\end{center}
\end{figure}

\subsection{Location \label{Locs}}

The observed features lie primarily in three belts within the mid-A Ring, between 126,750~km and 132,000~km from Saturn's center (\Fig{s} \ref{prop_locs}a, \ref{prop_locs}b).  We note that this is same region of the mid-A~Ring in which self-gravity wakes \citep{Hedman07} and the azimuthal brightness asymmetry \citep{Dones93,French07} are at their peaks, though the nature of these effects' connection (if any) with propeller abundance is unknown at present.  The two exceptions are feature 013-20-A, a resolved propeller just outside the Encke Gap at 134,082~km; and feature 028-30-A, an unresolved feature in the outer-A Ring (134,715~km) that is morphologically similar to other features seen in the Orbit~28 data.  The belts are divided by propeller-poor regions between 128,000~km and 128,500~km, and between 130,200~km and 131,300~km.  Although the propeller-rich regions are relatively quiescent (i.e., unperturbed by strong density waves driven at satellite resonances) compared to the inner-A and outer-A Ring regions, other mid-A~Ring regions are just as quiescent but lack propellers (\Fig{}~\ref{CompareImages}).  Many propeller-rich regions also contain smaller-scale clumpy texture that is also oriented in the azimuthal direction.  

The outer of the two gaps between propeller-rich belts is centered on the strongest spiral density wave \citep[see][and references therein]{soirings} in this part of the A~Ring --- the Janus/Epimetheus 5:4, which propagates outward from 130,702~km --- but the propeller-poor region extends far beyond the wave both inward and outward.  The lack of propellers in this wider ``halo'' centered on a strong density wave may be related to other changes reported in such ``halos,'' including smaller regolith-grain sizes and a reduction in self-gravity wakes \citep{Hedman07,NichSOI07}.  Similarly, the inner edge of the propeller-rich region may be governed by the ``halo'' of the Janus/Epimetheus 4:3 density wave (125,268~km), and the outer edge by that of the Mimas~5:3 density wave (132,298~km).  On the other hand, the Mimas~5:3 bending wave, propagating inward from 131,900~km, appears to have little influence.  

The smaller inner gap between propeller-rich belts contains three moderate-strength spiral density waves --- Pandora~7:6, Mimas~7:4, and Janus/Epimetheus~9:7.  Although these are not significantly stronger than several other waves due to Prometheus and Pandora that punctuate the propeller-rich regions (see below), it is possible that the combination of the three waves together strengthens their impact on the local propeller population.  

\citet{Propellers06} suggested that moderate-strength spiral density waves locally disrupt the propeller population by increasing collision velocities and/or disrupting the development of self-gravity wakes, but our results suggest that such an effect occurs at most in the main part of the wave only (an annulus $\sim 50$~km in width).  Several propellers in the current data set are seen both radially inward and outward of moderate-strength spiral density waves --- 032-44-A and 046-10-P are very close to the Prometheus~9:8 resonance location (radially inward of the main part of the wave), while 013-12-K/013-13-B and 028-34-X/028-35-L and 046-10-O are $\sim 100$~km outward of the same resonance; similarly, 046-11-D is $\sim 100$~km outward of the Janus/Epimetheus~9:7 resonance location, and 013-08-B is only 65~km outward of the Prometheus~12:11 resonance location \citep[see also][]{Sremcevic07}.  In any case, propellers are more difficult to observe in the main part of a wave, due to the strong fluctuations in brightness.  

\subsection{Abundance \label{Abundance}}
Seeking a minimally biased population with which to estimate the intrinsic abundance of propellers, we here consider only resolved propellers with $\Delta r > 0.8$~km.  For smaller values of $\Delta r$, the rollover seen in \Fig{}~\ref{prop_locs}f indicates that detectability becomes an important effect.  The only observing sequences that contain more than one such large object are Orbit~13 and Orbit~46, with nine apiece.  Both are complete radial scans of the rings $0.4^\circ$ and $0.5^\circ$ wide in longitude, respectively, leading us to estimate that 7,000~to 8,000 propellers with $\Delta r > 0.8$~km exist in the A~Ring.  This yields a surface number density for large propellers of $2 \times 10^{-6}$~km$^{-2}$  
averaged over the annulus from 126,000 to 132,000~km from Saturn's center, though some regions within that annulus are an order of magnitude more propeller-rich than others (\Fig{}~\ref{prop_locs}a, Section~\ref{Locs}).  

Feature 013-20-A is the only propeller seen in a complete \textit{azimuthal} scan, in this case of the region surrounding the Encke Gap (this appears in a different observing sequence than the Orbit~13 \textit{radial} scan).  Therefore, it is likely the only propeller of that size existing in the annulus of the rings covered by that scan (133,000 to 134,200~km), yielding a surface number density at that location of $10^{-9}$~km$^{-2}$.  
Although \Cassit{} has carried out other complete azimuthal scans at various locations in the rings, none of them to date equal the high resolution of the Orbit~13 scan, and thus we cannot draw further conclusions from them about propeller abundances elsewhere. 

The surface number density of smaller propellers can be estimated from the 4 examples seen in SOI images, in which they are quite well-resolved, yielding a value of $7 \times 10^{-4}$~km$^{-2}$ for propellers with $\Delta r > 0.25$~km \citep{Propellers06}.  Given that the SOI images were not even looking at the most propeller-rich regions, the surface number density at some locations could be several times higher than this number.  On the other hand, in \Fig{}~\ref{prop_locs}a, the SOI propellers fall into bins that have $\sim 60$\% more propellers than the average over all of the propeller-rich belt. 
Although \citeauthor{Propellers06} estimated a total of $10^7$ propellers of this size by assuming they were homogeneously distributed through the A~Ring (a number that may yet describe an original population that has perhaps been since eroded in most locations), we now know that the radial distribution of propellers is highly non-uniform.  Still, we can state that $\sim 60,000$ propellers with $\Delta r > 0.25$~km exist merely in the 100-km-wide annulus defined by the two SOI images. 

\begin{figure}[!t]
\begin{center}
\includegraphics[width=8cm,keepaspectratio=true]{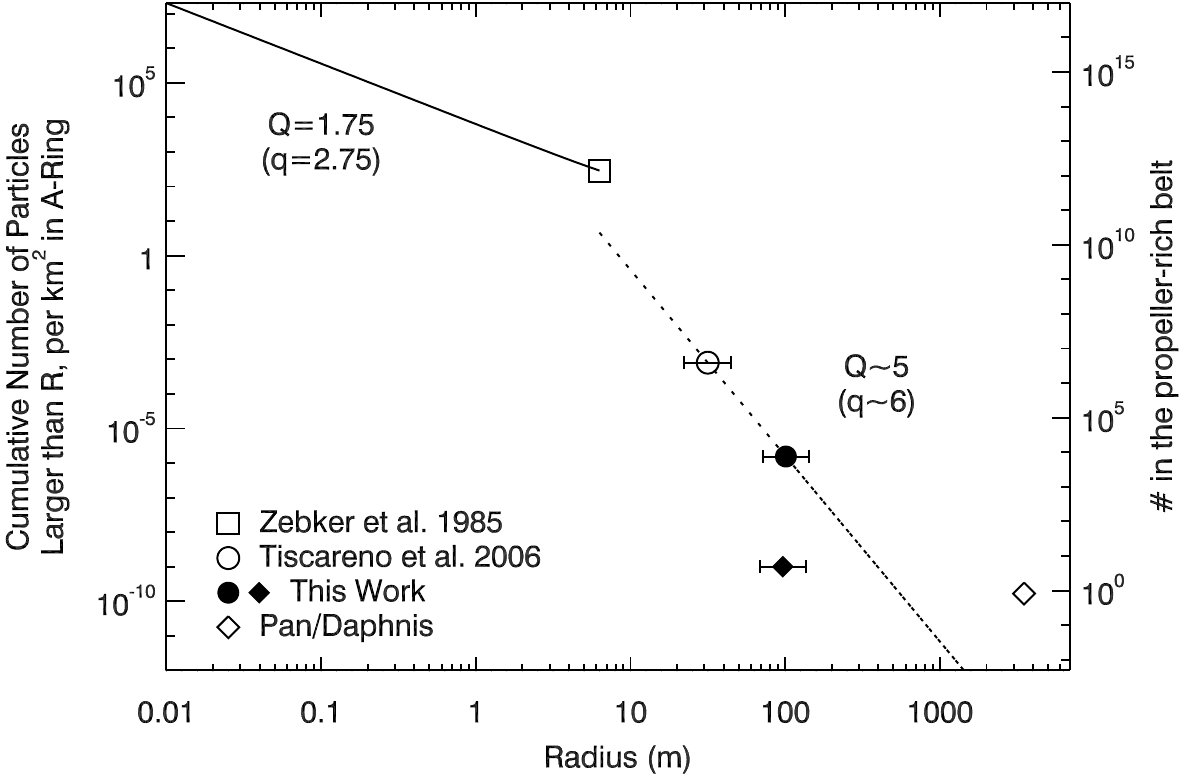}
\caption{Particle size distribution for Saturn's A Ring.  The solid line and open square denote the Voyager~RSS size distribution for the ring continuum \citep{Zebker85}, converted to integral format.  The open circle and the filled circle denote values for the propeller-rich belt, from \citet{Propellers06} and this work, respectively.  The error bars reflect the systematic error due to model-dependency in the conversion from $\Delta r$ to moonlet size (see Sections~\ref{MMass} and~\ref{Interpretation}); it is important to note that data points will slide along the error bars in concert, preserving their relative positions and thus the inferred power law.  The filled diamond is obtained from detection of one propeller near the Encke Gap with complete coverage of that annulus (\citeauthor{Sremcevic07} 2007; this work).  The open diamond is obtained from two known moons (Pan and Daphnis) of radius $\gtrsim 4$~km in the entire A~Ring.  \label{prop_powerlaw}}
\end{center}
\end{figure}

Comparing this small-propeller surface number density with the large-propeller surface number density derived above (and assuming that the size of the perturbing moonlet scales linearly with $\Delta r$) yields a steep particle-size distribution\fn{We refer to an integral size distribution, of the form $N(R) \propto R^{-Q}$, where $N$ is the number of particles per unit area with radius greater than $R$.} of $Q \sim 5.4$. 
Scaling the small-propeller density down by 60\%, to approximate the broader radial average of the large-propeller density, gives a slope only slightly gentler at $Q \sim 4.5$. 
Note that we do not include the Encke Gap data point in our fit, as its low value is attributable to the large radial variation in propeller abundance.  Our result is much shallower than the $Q \sim 8$~to~10 quoted by \citet{Sremcevic07}, but is closer to the power law inferred by \citet{Propellers06} to connect the Voyager small-particle size distribution to Pan and Daphnis.  In connecting different particle populations to one another, it is worth noting that considerable evolution in particle sizes may have occurred \citep{PorcoSci07}. 

\subsection{Dimensions \label{Dims}}
The total lengths and widths of observed features are shown in \Fig{s}~\ref{prop_locs}c and~\ref{prop_locs}d.  For the single-gaussian model (\Eqn{}~\ref{SingleGaussianModel}) these are simply $a$ and $b$, respectively.  For the primary model (\Eqn{}~\ref{DoubleGaussianModel}), however, the total length is $a+\Delta\ell$ and the total width is $b+\Delta r$ (\Fig{}~\ref{PropModel}).  These measured quantities depend partly on the image resolution, but still provide a measure of relative sizes and/or brightnesses.  Most features in our data set are between 4~and 12~km in length, and between 1.4~and 3.4~km in width.  However, the SOI features (shown in black) are much smaller, only $\sim 3$~km long and $\sim 0.5$~km wide, an effect probably attributable to the superior resolution of the SOI images (see below).  Furthermore, although one might expect unresolved features to be propellers that are too small to be resolved in the current images, their characteristic sizes in each observing sequence are only slightly smaller than resolved features in the same sequence.  This may indicate that apparent sizes are primarily determined by observational factors other than actual propeller sizes, or alternatively that many of the unresolved features are not propellers at all. 

\begin{figure}[!t]
\begin{center}
\includegraphics[width=16cm,keepaspectratio=true]{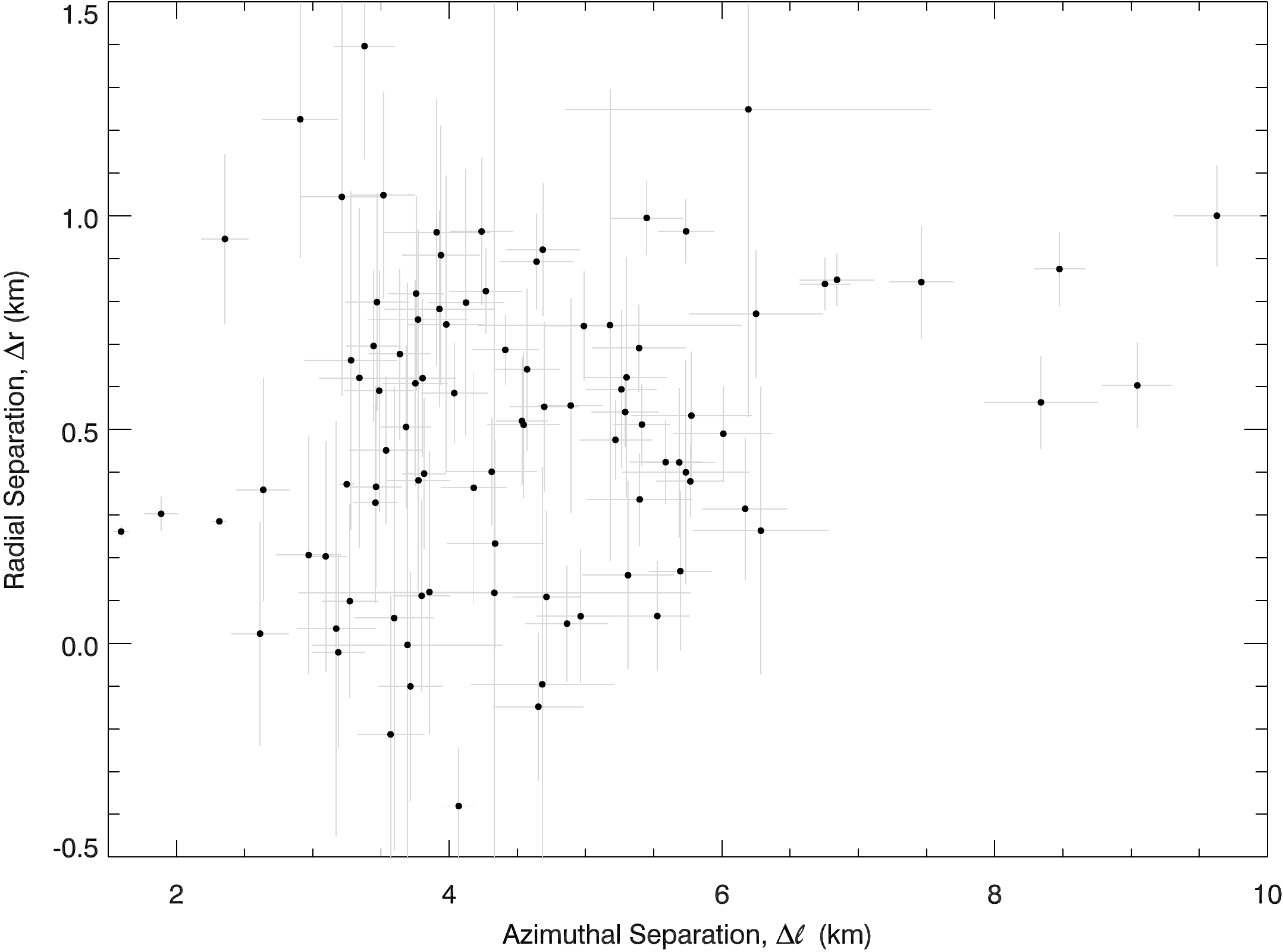}
\caption{Radial separation plotted versus azimuthal separation.   Three propellers --- 013-20-A, 028-36-B, and 031-47-A --- are not shown due to their large values of $\Delta\ell$.  \label{rad_v_length}}
\end{center}
\end{figure}

For resolved propellers, \Fig{s}~\ref{prop_locs}e and~\ref{prop_locs}f show the azimuthal and radial separations between the two lobe centers, $\Delta\ell$ and $\Delta r$, respectively.  Unlike the measurements discussed in the previous paragraph, these are ideally independent of image resolution.  Azimuthal separations are narrowly confined between 3~and 6~km, and radial separations are 1~km or less with a mean value of 0.5~km (note that a few $\Delta r$ values are $<0$, which would imply a reversed keplerian shear, but most of these have error bars that overlap zero).  Radial separation is the only measurement for which the very well-resolved SOI propellers are not the smallest propellers in our data set; it is also the measurement most closely related to moonlet size, leading to the conclusion that the SOI propellers' anomaly in other dimensional measurements merely reflects the improved resolution of the SOI images. 

We note that the sharp drop with increasing size in the number of large objects seen in \Fig{}~\ref{prop_locs}, for all four of the dimensions shown, is consistent with a very steep particle-size distribution (Section~\ref{Abundance}).  The lower-size cutoff, of course, is governed by detectability.  

We checked for variation of the various size parameters with location in the ring, as well as for variation in the frequency of resolved vs. unresolved features.  We found no clear evidence that any parameter other than the number of propellers varies with distance from Saturn.  

Figure~\ref{rad_v_length} plots the azimuthal and radial separations against each other, with error estimates fully shown.  Although \citet{Sremcevic07} claimed to notice a linear relationship between $\Delta\ell$ and $\Delta r$, we find little evidence of such a linear relationship in our larger sample size.  The Pearson correlation coefficient between $\Delta\ell$ and $\Delta r$ is only 0.24. 

Analytical theory indicates that the length of propeller-shaped gaps should scale as $(\Delta r)^3/\nu$ \citep{SSD02}, 
where $\nu$ is the local dynamical viscosity, as random non-keplerian velocities cause ring particles to fill in the propeller's gaps through diffusion.  This relation was confirmed in the numerical models of \citet{Seiss05}.  \citet{Sremcevic07} claim their failure to fit their data with a curve of the form $\Delta\ell \propto (\Delta r)^3$ is evidence that observed propeller-shaped features are due to moonlet wakes rather than propeller-shaped gaps (see Section~\ref{Propellers} for definitions, and Section~\ref{Interpretation} for further discussion).  However, these analyses neglect the effects of mutual gravity among ring particles, particularly the self-gravity wakes that are ubiquitous in this region of the A~Ring \citep{JT66,Salo95,Colwell06,Hedman07}.  More detailed simulations of propellers in the A~Ring's environment are ongoing \citep[e.g.,][]{LS07,LewisDDA07,Sremcevic07}, but have yet to put significant constraints on length scaling.  Furthermore, the local dynamical viscosity ($\nu$) in the propeller-rich region was recently shown to fluctuate significantly, as well as generally increasing by nearly an order of magnitude from 126,750 to 132,000~km from Saturn's center, as indicated by modeling of spiral density waves \citep{soirings}.  Thus, drawing robust conclusions from the length of propellers will be a complex task.  

\subsection{Brightness \label{Brightness}}
\begin{figure}[!t]
\begin{center}
\includegraphics[width=10cm,keepaspectratio=true]{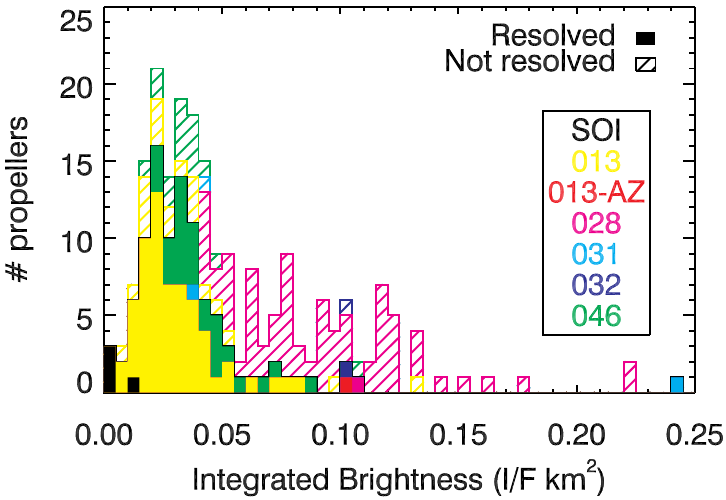}
\caption{Integrated brightness from propeller model fits.  Not shown is 028-33-C, with a value of 0.35.  \label{PropBrightness}}
\end{center}
\end{figure}

The integrated brightness above the background for modeled propellers is $\pi A_1 a b$ for the primary model (\Eqn{}~\ref{DoubleGaussianModel}) and $\pi A_1 a b / 2$ for the secondary model (\Eqn{}~\ref{SingleGaussianModel}).  This quantity is plotted in \Fig{}~\ref{PropBrightness}.  All features in our data set are brighter than their background (i.e., $A_1 > 0$), though dark features should be just as easily detectable.  Similar brightnesses are obtained for the two highest-resolution observing sequences:  Orbits~13 and~46.  Though these two sequences are similar in resolution (Table~\ref{ObserveInfo}), areal coverage, and number of large propellers detected (Section~\ref{Abundance}), they differ in that Orbit~13 is at high phase angle while Orbit~46 is at low phase angle (Table~\ref{ObserveInfo}).  We also note that Orbit~13 views the unlit side of the rings, Orbit~46 the lit side.  (A higher median brightness is obtained for the low-phase Orbit~28 data, but this is likely a result of dimmer objects being undetectable in the lower-resolution data, rather than a phase effect.)  The lack of significant brightening with increasing phase angle (comparing Orbit~13 to Orbit~46) indicates that the observed features are likely composed primarily of macroscopic particles, rather than dust. 

\subsection{Moonlet Mass and Size \label{MMass}}
Analytical theory indicates that the radial separation between the two lobes of a propeller-shaped gap ($\Delta r$) should be 4~times the Hill radius of the central moonlet (Section~\ref{Propellers}).  Unfortunately, detailed numerical simulations that take particle self-gravity into account have not directly addressed this issue, but preliminary indications are that the result $\Delta r \sim 4 r_H$ holds (D.~C.~Richardson, private communication, 2006).  On the other hand, the average separation between moonlet wakes (see \Fig{}~\ref{Illustration}) is larger, from $6 r_H$ to $10 r_H$; it is also more difficult to quantify (as moonlet wakes do not generally lie along the azimuthal direction) and varies with the parameters of a given simulation.  

We can thus infer properties of the embedded moonlet, if we assume we are observing propeller gaps (see Section~\ref{Interpretation}); in this case, \Fig{}~\ref{prop_locs}f indicates that the largest embedded moonlets ($\Delta r \lesssim 1.4$~km) have masses $\lesssim 3 \times 10^{13}$~g and diameters $D \lesssim 500$~m if their density is equal to the Roche critical density\fn{The Roche critical density, $\rho_R \sim 1.88 M_S / a^3$ for Saturn's mass $M_S$ and the moonlet's orbital radius $a$, is the density of an object whose physical size is the same as the size of its region of gravitational dominance.} of 0.5~g cm$^{-3}$ \citep{PorcoSci07}.  A lower limit on propeller size can be drawn from the conclusion of \citet{LS07} that propellers do not form if the embedded moonlet is not significantly larger than the largest particles that are abundant in the ring continuum.  Thus, a propeller-causing moonlet must have $D \gtrsim 50$~m, which, using the assumptions given in this paragraph, would result in $\Delta r \gtrsim 140$~m.  In our data set, only 3 of the 78 resolved propellers (4\%) have error bars that are entirely below this value; the discovery of a significant population of resolved propellers with $\Delta r \ll 140$~m would be evidence against interpreting observed propellers as gaps. 

\section{Interpretation of Propeller-shaped Features \label{Interpretation}}

For images of the unlit side of the rings (including the original SOI detections), some ambiguity exists in principle as to whether locally bright features are due to an increase or a decrease in local optical depth \citep{Propellers06}.  For propellers, this ambiguity is largely resolved by the further data reported here, including many images of the lit side of the rings, that indicate that propellers are generally local increases in the effective optical depth --- which is to say that the observed structures contain more material interacting with light than does the background.  From this one might conclude that what is being seen are the density enhancements (wakes) around the propeller, rather than the propeller-shaped gaps themselves (see Section~\ref{Propellers}).  

However, this inference confronts two major problems.  Firstly, the morphology of observed propellers strongly favors gaps over moonlet wakes: the bright bands in \Fig{}~\ref{PropExamples} appear like the open regions in \Fig{}~\ref{Illustration}.  Secondly, because a smaller central moonlet is required to produce a given observed value of $\Delta r$, if it refers to the separation between moonlet wakes, this interpretation results in more inferred central moonlets (12\%, rather than the 4\% reported above) that would not be significantly larger than the largest particles that are abundant in the ring continuum, a condition under which simulated propellers do not form according to some models \citep{LS07}.  Furthermore, in simulations presented by \citet{LewisDDA07}, propeller gaps appear under certain conditions without any significant associated moonlet wakes.  However, the converse is not true, and the question then arises: if it is wakes that are being observed, where are the gaps?

One possible resolution of this conundrum has been proposed by \citet{Sremcevic07}: that ring-particle regolith is temporarily liberated by the effects of the embedded moonlet, locally increasing the optical depth (but not the mass density) within the propeller until the material is re-accreted onto continuum ring particles.  

A role may also be played by a moonlet's varying effects on ring particles of various sizes.  Due to gravitational scattering by larger ring particles, smaller particles (with sizes between 1~and 10~cm) may already have eccentricities as large as that imparted by an encounter with the moonlet, and thus may be less affected by the moonlet.  These small particles interact more effectively with light than a large particle of the same total mass, but are too numerous to include in present numerical simulations; they may be at least partly reponsible for propeller gaps that are brighter than simulations predict.\fn{The authors are indebted to the anonymous reviewer for this idea.}  A similar effect may have been seen in the Keeler Gap, where brightness is enhanced in the wavy gap edges surrounding the moon Daphnis, even though numerical simulations indicate a decrease in particle density at such locations \citep{LewisDPS06}. 

We suggest, only briefly in this paper, a different way in which the \textit{effective} optical depth need not correlate exactly with the total amount of material present.  The A~Ring is not a homogeneous sheet of material, but rather consists of a matrix of self-gravity wakes that are nearly opaque, while only the low-$\tau$ spaces between the wakes are able to transmit light to an obsever on the unlit side of the rings \citep[see][]{Colwell06,Hedman07}.  A moonlet embedded in the ring may locally disrupt the self-gravity wakes, releasing more material into the optically active component even as the total amount of material is lessened in the propeller gap.  We hope to further develop this idea in a future paper.    

\acknowledgements{We thank M. Srem{\v c}evi{\'c}, J. Schmidt, and P. Nicholson for helpful discussions, and an anonymous reviewer for improvements to the manuscript.  We acknowledge funding by \Cassit{} and by NASA~PG\&G.  This paper is dedicated to FCT.}

\newpage
\appendix

\renewcommand{\theequation}{A\arabic{equation}}
\setcounter{equation}{0}
\renewcommand{\thefigure}{A\arabic{figure}}
\setcounter{figure}{0}
\renewcommand{\thetable}{A\arabic{table}}
\setcounter{table}{0}

\begin{figure}[!t]
\begin{center}
\includegraphics[width=14cm,keepaspectratio=true]{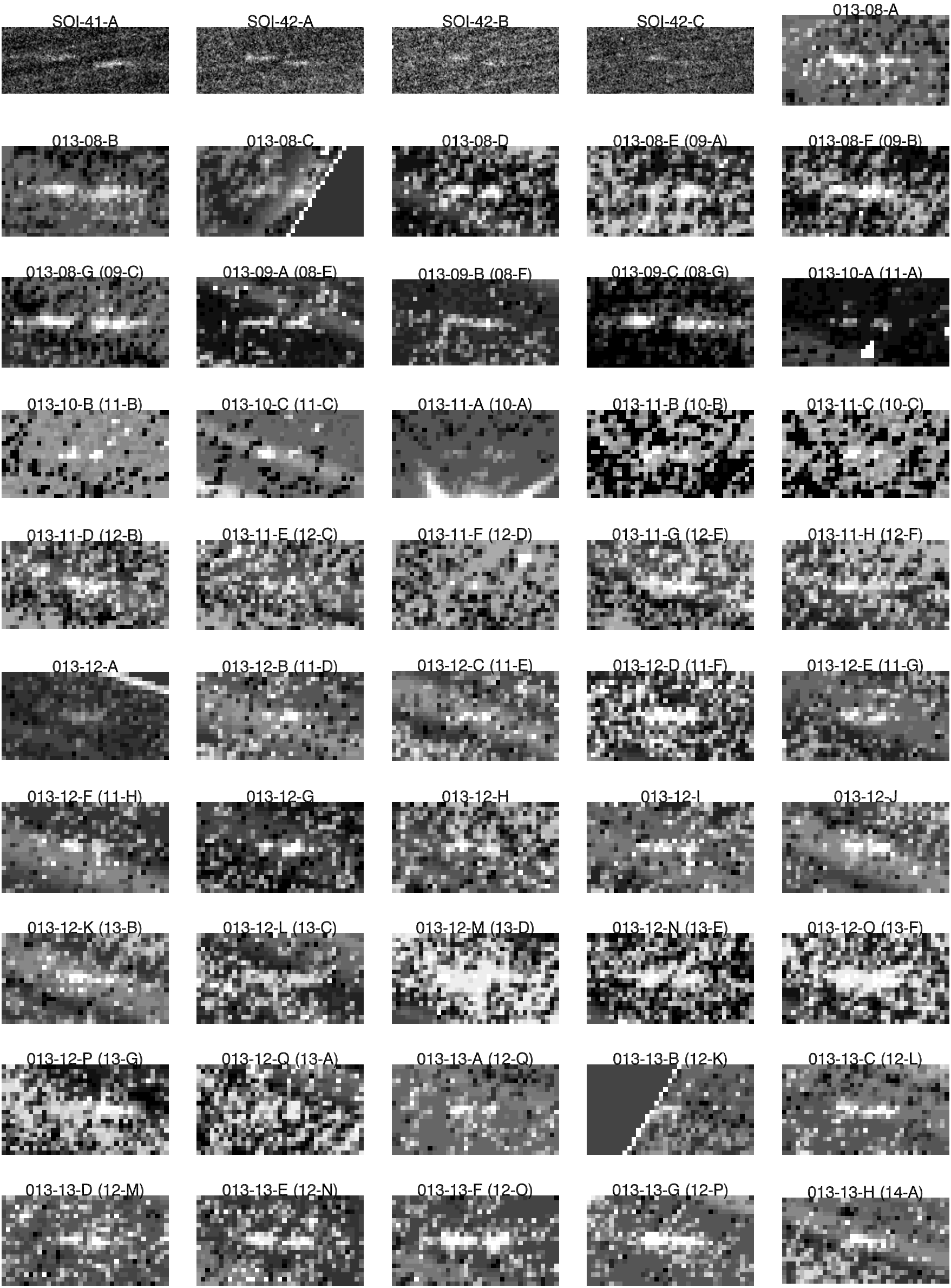}
\caption{Reprojected images centered on each of the features reported in our data set.  The direction towards Saturn is down, and the orbital direction is to the right.  The name of the feature (see Table~\ref{PropTable}) is given over each image.  If a feature's co-rotating location appears in a second image, the name of the matching feature is given in parenthesis.  \label{PropReproj}}
\end{center}
\end{figure}

\setcounter{figure}{0}
\begin{figure}[!t]
\begin{center}
\includegraphics[width=14cm,keepaspectratio=true]{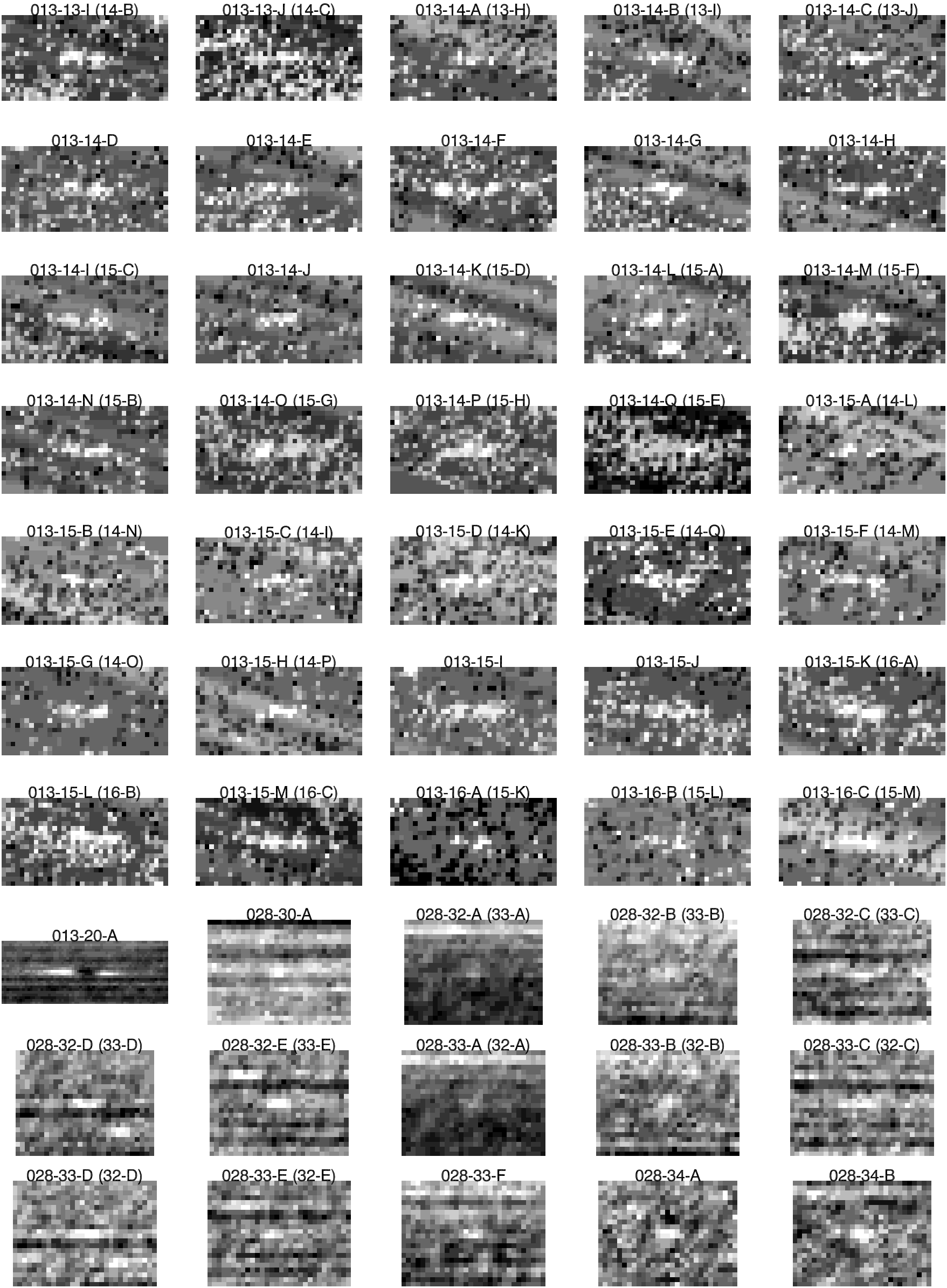}
\caption{continued}
\end{center}
\end{figure}

\setcounter{figure}{0}
\begin{figure}[!t]
\begin{center}
\includegraphics[width=14cm,keepaspectratio=true]{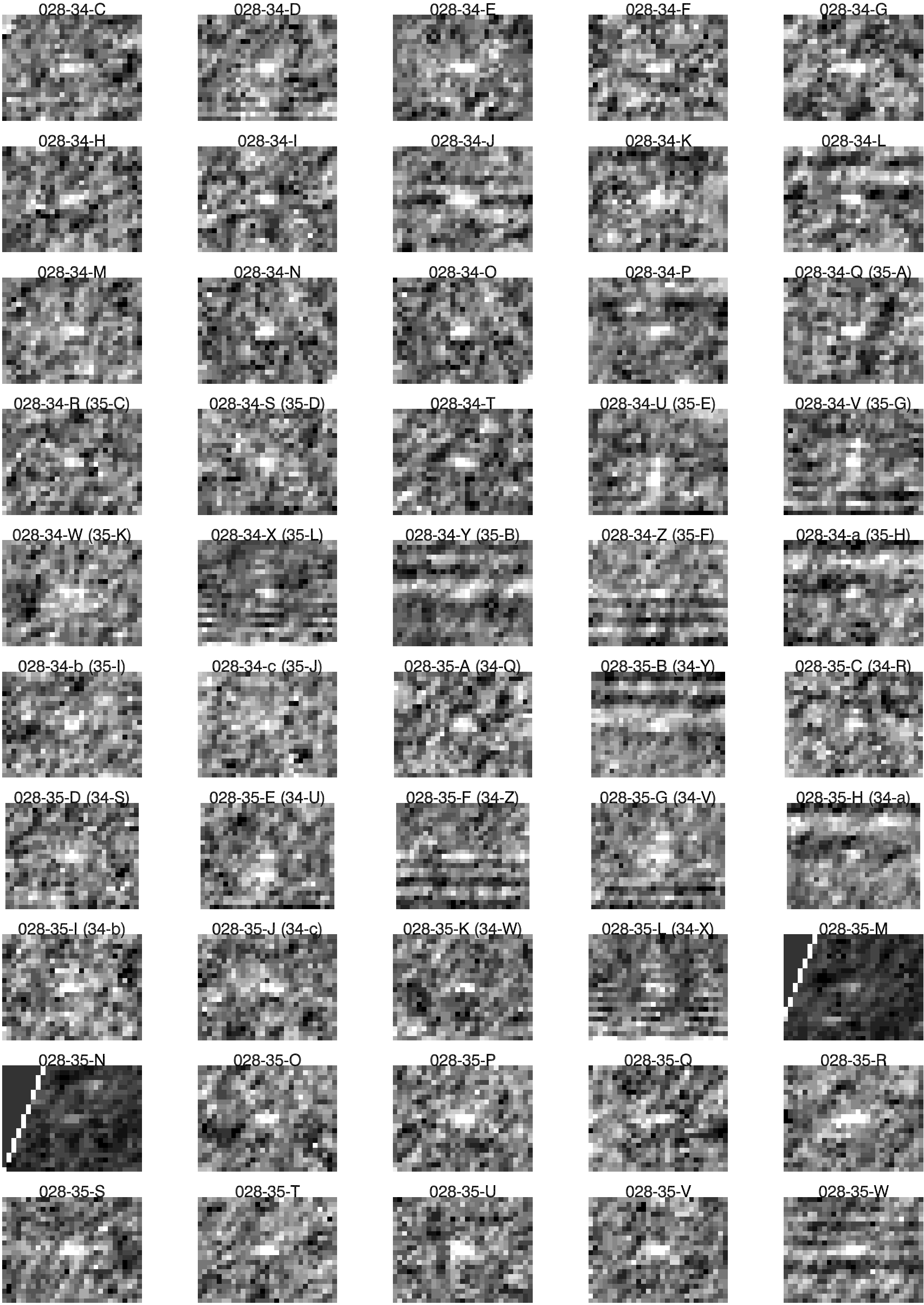}
\caption{continued}
\end{center}
\end{figure}

\setcounter{figure}{0}
\begin{figure}[!t]
\begin{center}
\includegraphics[width=14cm,keepaspectratio=true]{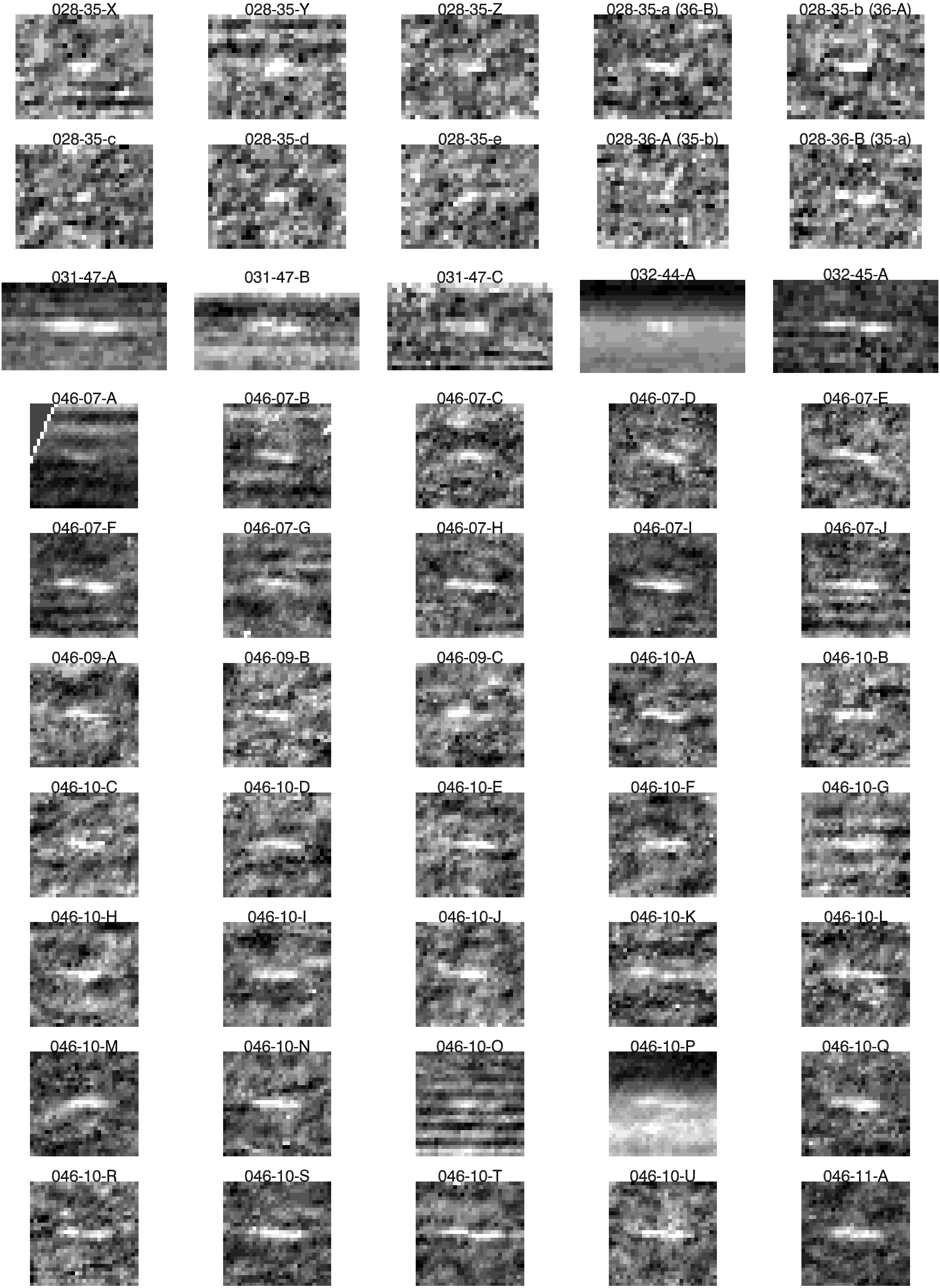}
\caption{continued}
\end{center}
\end{figure}

\begin{landscape}
\setcounter{figure}{0}
\begin{figure}[!t]
\begin{center}
\includegraphics[width=19cm,keepaspectratio=true]{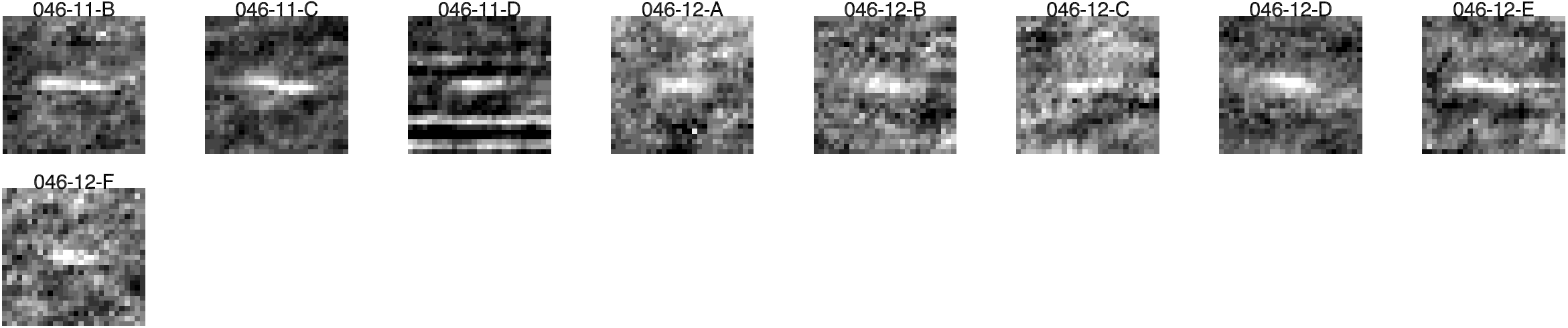}
\caption{continued}
\end{center}
\end{figure}

\begin{scriptsize}
\begin{longtable}[!t]{ l c c c c r @{$\pm$} l r @{$\pm$} l r @{$\pm$} l r @{$\pm$} l c }
\multicolumn{14}{l}
{\normalsize \tablename\ \thetable{}: Locations and fitted parameters for reported features} \\
\hline 
\hline
Name$^a$ & Image & [ line, sample ] & Radius (km)$^b$ & Longitude ($^\circ$)$^b$ & \multicolumn{2}{c}{$a$ (km)} & \multicolumn{2}{c}{$b$ (km)} & \multicolumn{2}{c}{$\Delta \ell$ (km)$^c$} & \multicolumn{2}{c}{$\Delta r$ (km)$^c$} & Match$^d$\\
\hline 
\endfirsthead
\multicolumn{14}{l}
{\normalsize \tablename\ \thetable{} --- continued from previous page} \\
\hline 
\hline
Name$^a$ & Image & [ line, sample ] & Radius (km)$^b$ & Longitude ($^\circ$)$^b$ & \multicolumn{2}{c}{$a$ (km)} & \multicolumn{2}{c}{$b$ (km)} & \multicolumn{2}{c}{$\Delta \ell$ (km)$^c$} & \multicolumn{2}{c}{$\Delta r$ (km)$^c$} & Match$^d$\\
\hline 
\endhead
\hline 
\multicolumn{14}{r}{Continued on next page} \\ 
\endfoot
\hline
\multicolumn{14}{l}{$^a$ Format is Orbit-Num-Letter, where Orbit and Num identify the image (see Table~\ref{ObserveInfo}), and Letter identifies the feature within the image.  In the case of more} \\
\multicolumn{14}{l}{than 26 features appearing in one image, lowercase letters are used beginning with ``a'' for the 27th feature.} \\ 
\multicolumn{14}{l}{$^b$ Radius is measured from Saturn's center.  Longitude is measured in an inertial coordinate system with the $z$-axis defined by Saturn's rotational pole at} \\ 
\multicolumn{14}{l}{epoch 1~January~2004 \citep{Jake06} and the $x$-axis defined by the ascending node of Saturn's equatorial plan on the J2000 equatorial plane.} \\ 
\multicolumn{14}{l}{$^c$ $\Delta \ell$ and $\Delta r$ are present if the fit to a propeller model (\Eqn{}~\ref{DoubleGaussianModel}) was successful.  When they are absent, the propeller structure is unresolved and the model fit}\\
\multicolumn{14}{l}{is to a simple gaussian peak (\Eqn{}~\ref{SingleGaussianModel}).}\\
\multicolumn{14}{l}{$^d$ If the co-rotating location appears in more than one image, the corresponding location is noted here.}\\
\endlastfoot
\label{PropTable} 
SOI-41-A & N1467347210 & [  182.6, 168.4 ] & 129498.39 & 324.78855 &  1.582 &  0.054 &  0.239 &  0.008 &  3.249 &  0.053 &  0.372 &  0.008 & \\
SOI-42-A & N1467347249 & [   82.8, 658.3 ] & 130099.86 & 325.11124 &  1.118 &  0.056 &  0.168 &  0.008 &  2.315 &  0.056 &  0.285 &  0.008 & \\
SOI-42-B &      "      & [  503.1, 500.1 ] & 130119.47 & 325.10610 &  0.742 &  0.123 &  0.240 &  0.040 &  1.888 &  0.123 &  0.303 &  0.040 & \\
SOI-42-C &      "      & [  664.2, 470.4 ] & 130127.26 & 325.10481 &  0.797 &  0.060 &  0.179 &  0.014 &  1.595 &  0.060 &  0.261 &  0.013 & \\
\hline
013-08-A & N1503229987 & [  175.9, 291.2 ] & 131941.89 & 280.83982 &  3.730 &  0.315 &  1.777 &  0.158 &  6.170 &  0.312 &  0.315 &  0.166 & \\
013-08-B &      "      & [  447.0, 238.1 ] & 131655.92 & 280.80526 &  4.479 &  0.241 &  2.265 &  0.133 &  7.462 &  0.238 &  0.845 &  0.132 & \\
013-08-C &      "      & [  528.3,   8.6 ] & 131505.34 & 280.84144 &  2.525 &  0.226 &  2.517 &  0.252 &  4.892 &  0.238 &  0.556 &  0.251 & \\
013-08-D &      "      & [  536.2,  41.5 ] & 131507.37 & 280.83289 &  1.937 &  0.184 &  1.495 &  0.155 &  4.534 &  0.191 &  0.520 &  0.151 & \\
013-08-E &      "      & [  703.7, 752.9 ] & 131555.40 & 280.64876 &  2.144 &  0.233 &  1.687 &  0.194 &  4.713 &  0.247 &  0.108 &  0.201 & 013-09-A\\
013-08-F &      "      & [  711.4, 225.4 ] & 131387.49 & 280.76265 &  7.530 &  0.656 &  2.296 &  0.201 & \multicolumn{2}{c}{} & \multicolumn{2}{c}{}  & 013-09-B\\
013-08-G &      "      & [  757.3, 824.8 ] & 131523.67 & 280.62376 &  5.017 &  0.250 &  1.825 &  0.102 &  9.046 &  0.256 &  0.603 &  0.100 & 013-09-C\\
013-09-A & N1503230047 & [   97.1, 909.3 ] & 131554.98 & 281.09351 &  3.367 &  0.249 &  1.582 &  0.122 &  5.527 &  0.241 &  0.064 &  0.129 & 013-08-E\\
013-09-B &      "      & [  106.6, 380.7 ] & 131387.34 & 281.20796 &  5.884 &  0.603 &  1.494 &  0.186 & \multicolumn{2}{c}{} & \multicolumn{2}{c}{}  & 013-08-F\\
013-09-C &      "      & [  150.6, 981.3 ] & 131523.23 & 281.06864 &  4.465 &  0.183 &  1.945 &  0.090 &  8.475 &  0.189 &  0.875 &  0.087 & 013-08-G\\
013-10-A & N1503230107 & [  829.1, 245.4 ] & 129968.08 & 281.49599 &  1.462 &  1.434 &  0.716 & 40.274 &  4.332 &  1.437 &  0.118 & 13.306 & 013-11-A\\
013-10-B &      "      & [  903.1, 100.1 ] & 129848.64 & 281.51619 &  0.991 &  0.144 &  1.768 &  0.282 &  3.095 &  0.154 &  0.203 &  0.270 & 013-11-B\\
013-10-C &      "      & [  983.9, 493.5 ] & 129884.53 & 281.41333 &  1.690 &  0.180 &  2.002 &  0.237 &  3.464 &  0.191 &  0.366 &  0.237 & 013-11-C\\
013-11-A & N1503230167 & [  238.1, 383.2 ] & 129967.83 & 281.94908 &  2.309 &  0.309 &  1.479 &  0.230 &  5.687 &  0.268 &  0.423 &  0.175 & 013-10-A\\
013-11-B &      "      & [  312.8, 237.9 ] & 129848.46 & 281.96963 &  8.043 &  1.795 &  2.640 &  0.603 & \multicolumn{2}{c}{} & \multicolumn{2}{c}{}  & 013-10-B\\
013-11-C &      "      & [  391.7, 630.6 ] & 129884.59 & 281.86712 &  5.233 &  1.156 &  3.173 &  0.702 & \multicolumn{2}{c}{} & \multicolumn{2}{c}{}  & 013-10-C\\
013-11-D &      "      & [  676.5, 238.4 ] & 129475.54 & 281.90843 &  2.149 &  0.329 &  2.135 &  0.316 &  2.908 &  0.280 &  1.226 &  0.325 & 013-12-B\\
013-11-E &      "      & [  754.4, 348.2 ] & 129428.21 & 281.87010 &  1.327 &  0.280 &  2.188 &  0.493 &  3.172 &  0.289 &  0.034 &  0.485 & 013-12-C\\
013-11-F &      "      & [  814.4, 121.2 ] & 129298.49 & 281.91195 &  1.621 &  0.326 &  1.811 &  0.408 &  3.280 &  0.340 &  0.662 &  0.396 & 013-12-D\\
013-11-G &      "      & [  832.1, 304.9 ] & 129335.14 & 281.86681 &  3.148 &  0.509 &  2.016 &  0.284 &  3.908 &  0.393 &  0.961 &  0.311 & 013-12-E\\
013-11-H &      "      & [  863.1,  88.7 ] & 129238.52 & 281.91113 &  7.598 &  1.233 &  1.884 &  0.312 & \multicolumn{2}{c}{} & \multicolumn{2}{c}{}  & 013-12-F\\
013-12-A & N1503230227 & [   10.9, 127.7 ] & 129488.06 & 282.43255 &  2.199 &  0.305 &  2.101 &  0.217 &  2.614 &  0.213 &  0.023 &  0.262 & \\
013-12-B &      "      & [   91.5, 368.9 ] & 129476.27 & 282.36389 &  1.727 &  0.194 &  1.791 &  0.232 &  3.271 &  0.206 &  0.099 &  0.227 & 013-11-D\\
013-12-C &      "      & [  169.8, 477.5 ] & 129427.87 & 282.32597 &  1.527 &  0.195 &  1.827 &  0.256 &  2.637 &  0.199 &  0.359 &  0.261 & 013-11-E\\
013-12-D &      "      & [  229.6, 250.9 ] & 129299.55 & 282.36826 &  2.938 &  0.470 &  2.155 &  0.301 &  3.855 &  0.368 &  0.120 &  0.331 & 013-11-F\\
013-12-E &      "      & [  247.8, 436.4 ] & 129335.56 & 282.32251 &  2.178 &  0.236 &  2.000 &  0.235 &  3.753 &  0.240 &  0.608 &  0.241 & 013-11-G\\
013-12-F &      "      & [  279.0, 215.9 ] & 129238.42 & 282.36818 &  1.714 &  0.162 &  2.459 &  0.264 &  3.459 &  0.163 &  0.329 &  0.234 & 013-11-H\\
013-12-G &      "      & [  439.8, 356.1 ] & 129113.85 & 282.30907 &  1.886 &  0.174 &  1.839 &  0.190 &  3.684 &  0.185 &  0.506 &  0.190 & \\
013-12-H &      "      & [  467.3, 579.3 ] & 129151.57 & 282.25289 &  1.878 &  0.268 &  2.036 &  0.314 &  4.123 &  0.281 &  0.797 &  0.313 & \\
013-12-I &      "      & [  550.5, 727.0 ] & 129109.51 & 282.20477 &  2.944 &  0.393 &  1.747 &  0.234 &  4.337 &  0.354 &  0.233 &  0.245 & \\
013-12-J &      "      & [  665.6, 997.0 ] & 129070.95 & 282.12293 &  2.692 &  0.269 &  2.526 &  0.233 &  3.775 &  0.230 &  0.381 &  0.259 & \\
013-12-K &      "      & [  666.5, 899.5 ] & 129040.94 & 282.14529 &  6.369 &  0.657 &  3.048 &  0.315 & \multicolumn{2}{c}{} & \multicolumn{2}{c}{}  & 013-13-B\\
013-12-L &      "      & [  715.7,  42.1 ] & 128735.04 & 282.33581 &  3.630 &  0.489 &  1.792 &  0.254 &  5.735 &  0.469 &  0.400 &  0.262 & 013-13-C\\
013-12-M &      "      & [  765.2, 672.7 ] & 128871.12 & 282.18101 &  9.900 &  1.135 &  6.016 &  0.718 & \multicolumn{2}{c}{} & \multicolumn{2}{c}{}  & 013-13-D\\
013-12-N &      "      & [  808.2, 693.1 ] & 128832.60 & 282.16896 &  2.950 &  0.293 &  2.552 &  0.281 &  5.301 &  0.302 &  0.622 &  0.282 & 013-13-E\\
013-12-O &      "      & [  808.9, 754.9 ] & 128850.36 & 282.15453 &  6.130 &  1.465 &  3.137 &  0.343 &  5.179 &  0.965 &  0.744 &  0.552 & 013-13-F\\
013-12-P &      "      & [  836.4, 635.6 ] & 128786.16 & 282.17752 &  4.799 &  0.637 &  2.606 &  0.296 &  6.286 &  0.501 &  0.263 &  0.337 & 013-13-G\\
013-12-Q &      "      & [  642.1, 395.0 ] & 128916.06 & 282.26620 &  1.478 &  0.279 &  2.742 &  0.550 &  3.597 &  0.289 &  0.059 &  0.543 & 013-13-A\\
013-13-A & N1503230287 & [   63.9, 517.1 ] & 128915.44 & 282.72490 &  1.904 &  0.195 &  1.954 &  0.226 &  3.798 &  0.208 &  0.111 &  0.225 & 013-12-Q\\
013-13-B &      "      & [   85.1,1019.0 ] & 129040.29 & 282.60463 &  2.060 &  0.274 &  1.213 &  0.204 & \multicolumn{2}{c}{} & \multicolumn{2}{c}{}  & 013-12-K\\
013-13-C &      "      & [  138.1, 162.7 ] & 128734.99 & 282.79551 &  2.749 &  0.246 &  1.924 &  0.184 &  4.571 &  0.245 &  0.641 &  0.190 & 013-12-L\\
013-13-D &      "      & [  187.2, 798.7 ] & 128870.47 & 282.63898 &  2.055 &  0.220 &  1.656 &  0.188 &  3.639 &  0.227 &  0.676 &  0.199 & 013-12-M\\
013-13-E &      "      & [  229.0, 815.2 ] & 128832.02 & 282.62821 &  2.710 &  0.248 &  1.828 &  0.185 &  5.263 &  0.261 &  0.594 &  0.186 & 013-12-N\\
013-13-F &      "      & [  229.4, 878.1 ] & 128850.08 & 282.61346 &  3.323 &  0.229 &  2.432 &  0.183 &  5.695 &  0.231 &  0.169 &  0.186 & 013-12-O\\
013-13-G &      "      & [  257.2, 756.2 ] & 128785.52 & 282.63729 &  3.598 &  0.317 &  2.293 &  0.174 &  4.697 &  0.251 &  0.553 &  0.198 & 013-12-P\\
013-13-H &      "      & [  906.2, 252.5 ] & 127959.70 & 282.64734 &  4.080 &  0.484 &  1.969 &  0.229 & \multicolumn{2}{c}{} & \multicolumn{2}{c}{}  & 013-14-A\\
013-13-I &      "      & [  947.4, 406.9 ] & 127962.23 & 282.60392 &  2.028 &  0.230 &  2.031 &  0.253 &  3.471 &  0.233 &  0.798 &  0.254 & 013-14-B\\
013-13-J &      "      & [  953.1, 477.4 ] & 127977.19 & 282.58630 &  6.716 &  1.279 &  1.562 &  0.244 & \multicolumn{2}{c}{} & \multicolumn{2}{c}{}  & 013-14-C\\
013-14-A & N1503230347 & [  333.9, 364.7 ] & 127960.12 & 283.11179 &  5.196 &  0.980 &  2.169 &  0.409 & \multicolumn{2}{c}{} & \multicolumn{2}{c}{}  & 013-13-H\\
013-14-B &      "      & [  375.5, 521.7 ] & 127962.37 & 283.06766 &  2.240 &  0.269 &  2.197 &  0.272 &  3.487 &  0.254 &  0.591 &  0.283 & 013-13-I\\
013-14-C &      "      & [  381.5, 594.7 ] & 127977.31 & 283.04932 &  3.384 &  0.538 &  1.579 &  0.217 & \multicolumn{2}{c}{} & \multicolumn{2}{c}{}  & 013-13-J\\
013-14-D &      "      & [  423.1,  40.3 ] & 127771.85 & 283.17448 &  2.008 &  0.196 &  2.097 &  0.228 &  3.758 &  0.205 &  0.818 &  0.227 & \\
013-14-E &      "      & [  454.3, 305.2 ] & 127816.38 & 283.10625 &  1.684 &  0.232 &  1.760 &  0.252 &  2.970 &  0.240 &  0.206 &  0.278 & \\
013-14-F &      "      & [  466.7, 143.7 ] & 127756.16 & 283.14270 &  3.556 &  0.377 &  2.033 &  0.212 &  5.312 &  0.335 &  0.159 &  0.220 & \\
013-14-G &      "      & [  528.3, 539.2 ] & 127807.00 & 283.03831 &  1.637 &  0.193 &  1.680 &  0.220 &  2.356 &  0.173 &  0.946 &  0.199 & \\
013-14-H &      "      & [  588.6, 984.8 ] & 127874.48 & 282.92229 &  2.550 &  0.260 &  1.423 &  0.167 &  4.546 &  0.267 &  0.511 &  0.171 & \\
013-14-I &      "      & [  605.7, 693.4 ] & 127770.89 & 282.98874 &  2.498 &  0.242 &  2.520 &  0.262 &  4.180 &  0.241 &  0.364 &  0.271 & 013-15-C\\
013-14-J &      "      & [  641.5, 969.2 ] & 127814.33 & 282.91714 &  4.340 &  0.440 &  1.971 &  0.196 & \multicolumn{2}{c}{} & \multicolumn{2}{c}{}  & \\
013-14-K &      "      & [  716.3,  37.8 ] & 127462.01 & 283.12710 &  2.680 &  0.240 &  1.815 &  0.181 &  4.238 &  0.231 &  0.963 &  0.172 & 013-15-D\\
013-14-L &      "      & [  830.8, 707.4 ] & 127537.82 & 282.94785 &  1.426 &  0.682 &  1.680 &  0.761 &  3.695 &  0.696 & -0.004 &  0.846 & 013-15-A\\
013-14-M &      "      & [  916.5, 784.0 ] & 127469.92 & 282.91513 &  2.542 &  0.267 &  2.439 &  0.248 &  3.716 &  0.237 & -0.101 &  0.268 & 013-15-F\\
013-14-N &      "      & [  939.4, 273.6 ] & 127295.11 & 283.03362 &  2.601 &  0.272 &  1.680 &  0.178 &  3.805 &  0.245 &  0.620 &  0.185 & 013-15-B\\
013-14-O &      "      & [ 1000.7, 769.3 ] & 127376.50 & 282.90448 &  1.500 &  0.111 &  1.818 &  0.136 &  4.070 &  0.112 & -0.381 &  0.137 & 013-15-G\\
013-14-P &      "      & [ 1014.6,  46.9 ] & 127148.62 & 283.07558 &  5.202 &  0.555 &  2.554 &  0.272 & \multicolumn{2}{c}{} & \multicolumn{2}{c}{}  & 013-15-H\\
013-14-Q &      "      & [  888.8, 310.0 ] & 127359.34 & 283.03331 & 12.422 &  1.107 &  4.911 &  0.413 & \multicolumn{2}{c}{} & \multicolumn{2}{c}{}  & 013-15-E\\
013-15-A & N1503230407 & [  264.2, 814.4 ] & 127537.29 & 283.41408 &  1.752 &  0.233 &  2.242 &  0.336 &  3.570 &  0.244 & -0.213 &  0.326 & 013-14-L\\
013-15-B &      "      & [  374.8, 379.4 ] & 127294.44 & 283.50085 &  1.914 &  0.310 &  0.731 & 19.560 &  3.213 &  0.310 &  1.044 &  0.465 & 013-14-N\\
013-15-C &      "      & [   40.1, 804.4 ] & 127769.95 & 283.45290 &  4.669 &  0.712 &  1.654 &  0.228 & \multicolumn{2}{c}{} & \multicolumn{2}{c}{}  & 013-14-I\\
013-15-D &      "      & [  151.9, 143.1 ] & 127461.58 & 283.59372 &  1.728 &  0.186 &  1.811 &  0.206 &  3.187 &  0.196 & -0.021 &  0.224 & 013-14-K\\
013-15-E &      "      & [  323.5, 413.1 ] & 127358.47 & 283.50103 &  2.692 &  0.308 &  2.889 &  0.327 &  3.979 &  0.279 &  0.746 &  0.348 & 013-14-Q\\
013-15-F &      "      & [  349.4, 890.3 ] & 127469.55 & 283.38189 &  1.881 &  0.194 &  1.541 &  0.197 &  3.446 &  0.202 &  0.695 &  0.176 & 013-14-M\\
013-15-G &      "      & [  431.3, 868.6 ] & 127376.77 & 283.37366 &  2.003 &  0.153 &  2.060 &  0.175 &  3.816 &  0.161 &  0.397 &  0.177 & 013-14-O\\
013-15-H &      "      & [  450.2, 151.1 ] & 127148.36 & 283.54380 &  4.824 &  0.438 &  2.450 &  0.222 & \multicolumn{2}{c}{} & \multicolumn{2}{c}{}  & 013-14-P\\
013-15-I &      "      & [  472.5, 160.6 ] & 127127.48 & 283.53789 &  3.025 &  0.561 &  2.533 &  0.477 &  4.683 &  0.528 & -0.096 &  0.508 & \\
013-15-J &      "      & [  472.5, 747.9 ] & 127298.16 & 283.39597 &  7.202 &  0.853 &  2.480 &  0.294 & \multicolumn{2}{c}{} & \multicolumn{2}{c}{}  & \\
013-15-K &      "      & [  670.2, 899.9 ] & 127133.10 & 283.32667 &  2.504 &  0.268 &  2.505 &  0.259 &  3.380 &  0.227 &  1.397 &  0.266 & 013-16-A\\
013-15-L &      "      & [  780.7, 741.6 ] & 126969.44 & 283.34665 &  2.869 &  0.342 &  2.543 &  0.308 &  3.940 &  0.285 &  0.908 &  0.304 & 013-16-B\\
013-15-M &      "      & [  913.2, 547.2 ] & 126771.38 & 283.37185 &  6.216 &  0.633 &  1.944 &  0.193 & \multicolumn{2}{c}{} & \multicolumn{2}{c}{}  & 013-16-C\\
013-16-A & N1503230467 & [  110.0,1001.9 ] & 127133.87 & 283.79496 &  1.652 &  0.282 &  2.098 &  0.398 &  3.342 &  0.299 &  0.621 &  0.397 & 013-15-K\\
013-16-B &      "      & [  221.0, 842.1 ] & 126970.41 & 283.81589 &  1.880 &  0.225 &  1.859 &  0.255 &  3.518 &  0.234 &  1.048 &  0.242 & 013-15-L\\
013-16-C &      "      & [  354.2, 643.8 ] & 126771.77 & 283.84276 &  7.170 &  0.633 &  2.695 &  0.239 & \multicolumn{2}{c}{} & \multicolumn{2}{c}{}  & 013-15-M\\
\hline
013-20-A & N1503243458 & [  221.7, 426.0 ] & 134082.40 & 327.52251 &  8.875 &  0.370 &  2.015 &  0.086 & 22.092 &  0.379 &  0.765 &  0.089 & \\
\hline
028-30-A & N1536497993 & [  771.6, 190.2 ] & 134714.95 &  67.26860 &  4.942 &  0.878 &  0.965 & 13.227 & \multicolumn{2}{c}{} & \multicolumn{2}{c}{}  & \\
028-32-A & N1536498176 & [  944.4,  31.4 ] & 131733.84 &  67.17195 & 10.094 &  1.395 &  3.047 &  0.451 & \multicolumn{2}{c}{} & \multicolumn{2}{c}{}  & 028-33-A\\
028-32-B &      "      & [  956.0,  91.4 ] & 131722.23 &  67.23078 & 11.159 &  1.889 &  3.278 &  0.550 & \multicolumn{2}{c}{} & \multicolumn{2}{c}{}  & 028-33-B\\
028-32-C &      "      & [  976.8, 193.4 ] & 131701.40 &  67.33038 & 19.275 &  3.301 &  2.430 &  0.354 & \multicolumn{2}{c}{} & \multicolumn{2}{c}{}  & 028-33-C\\
028-32-D &      "      & [  979.7, 299.8 ] & 131709.87 &  67.44200 & 11.327 &  1.624 &  0.992 &  5.944 & \multicolumn{2}{c}{} & \multicolumn{2}{c}{}  & 028-33-D\\
028-32-E &      "      & [  986.0, 310.4 ] & 131701.12 &  67.45054 & 17.404 &  1.979 &  3.416 &  0.362 & \multicolumn{2}{c}{} & \multicolumn{2}{c}{}  & 028-33-E\\
028-33-A & N1536498268 & [  108.4, 281.8 ] & 131736.49 &  67.85738 &  6.866 &  1.061 &  4.519 &  0.773 & \multicolumn{2}{c}{} & \multicolumn{2}{c}{}  & 028-32-A\\
028-33-B &      "      & [  120.1, 342.7 ] & 131725.74 &  67.91652 &  7.836 &  1.092 &  3.637 &  0.505 & \multicolumn{2}{c}{} & \multicolumn{2}{c}{}  & 028-32-B\\
028-33-C &      "      & [  142.5, 446.7 ] & 131704.08 &  68.01648 & 26.171 &  2.509 &  3.904 &  0.329 & \multicolumn{2}{c}{} & \multicolumn{2}{c}{}  & 028-32-C\\
028-33-D &      "      & [  146.4, 554.8 ] & 131712.71 &  68.12823 & 26.428 &  3.809 &  0.929 &  3.257 & \multicolumn{2}{c}{} & \multicolumn{2}{c}{}  & 028-32-D\\
028-33-E &      "      & [  153.0, 564.5 ] & 131703.52 &  68.13562 &  8.360 &  1.672 &  1.768 &  0.628 & \multicolumn{2}{c}{} & \multicolumn{2}{c}{}  & 028-32-E\\
028-33-F &      "      & [  247.6, 694.5 ] & 131572.57 &  68.23361 &  9.694 &  1.363 &  2.889 &  0.409 & \multicolumn{2}{c}{} & \multicolumn{2}{c}{}  & \\
028-34-A & N1536498361 & [  398.3, 406.8 ] & 129932.20 &  67.95499 &  7.920 &  1.115 &  2.497 &  0.372 & \multicolumn{2}{c}{} & \multicolumn{2}{c}{}  & \\
028-34-B &      "      & [  404.2,  82.4 ] & 129882.86 &  67.60592 &  8.811 &  1.118 &  2.917 &  0.374 & \multicolumn{2}{c}{} & \multicolumn{2}{c}{}  & \\
028-34-C &      "      & [  417.7, 585.0 ] & 129925.55 &  68.13655 & 11.341 &  1.626 &  2.602 &  0.386 & \multicolumn{2}{c}{} & \multicolumn{2}{c}{}  & \\
028-34-D &      "      & [  465.0, 668.2 ] & 129862.24 &  68.20541 &  7.852 &  1.042 &  3.206 &  0.426 & \multicolumn{2}{c}{} & \multicolumn{2}{c}{}  & \\
028-34-E &      "      & [  543.0, 166.0 ] & 129673.20 &  67.63668 & 11.238 &  1.354 &  2.773 &  0.339 & \multicolumn{2}{c}{} & \multicolumn{2}{c}{}  & \\
028-34-F &      "      & [  564.7, 535.0 ] & 129685.87 &  68.02235 &  9.127 &  1.378 &  2.142 &  0.400 & \multicolumn{2}{c}{} & \multicolumn{2}{c}{}  & \\
028-34-G &      "      & [  566.0, 482.4 ] & 129676.76 &  67.96574 & 14.801 &  1.841 &  2.801 &  0.354 & \multicolumn{2}{c}{} & \multicolumn{2}{c}{}  & \\
028-34-H &      "      & [  581.2, 755.2 ] & 129690.67 &  68.25001 &  8.450 &  1.592 &  2.801 &  0.532 & \multicolumn{2}{c}{} & \multicolumn{2}{c}{}  & \\
028-34-I &      "      & [  581.8, 460.8 ] & 129648.92 &  67.93609 &  5.994 &  0.949 &  1.927 &  0.467 & \multicolumn{2}{c}{} & \multicolumn{2}{c}{}  & \\
028-34-J &      "      & [  613.9, 160.9 ] & 129560.42 &  67.60115 &  7.929 &  1.008 &  3.717 &  0.474 & \multicolumn{2}{c}{} & \multicolumn{2}{c}{}  & \\
028-34-K &      "      & [  650.1, 106.0 ] & 129496.75 &  67.52663 & 11.387 &  1.409 &  4.162 &  0.517 & \multicolumn{2}{c}{} & \multicolumn{2}{c}{}  & \\
028-34-L &      "      & [  692.4, 711.2 ] & 129507.93 &  68.15721 &  5.558 &  0.930 &  1.669 &  0.639 & \multicolumn{2}{c}{} & \multicolumn{2}{c}{}  & \\
028-34-M &      "      & [  706.5, 755.1 ] & 129491.90 &  68.19814 &  9.067 &  1.406 &  2.424 &  0.410 & \multicolumn{2}{c}{} & \multicolumn{2}{c}{}  & \\
028-34-N &      "      & [  732.7, 765.6 ] & 129451.89 &  68.19850 &  5.921 &  0.723 &  2.609 &  0.327 & \multicolumn{2}{c}{} & \multicolumn{2}{c}{}  & \\
028-34-O &      "      & [  732.7, 765.6 ] & 129451.90 &  68.19850 &  5.930 &  0.725 &  2.639 &  0.333 & \multicolumn{2}{c}{} & \multicolumn{2}{c}{}  & \\
028-34-P &      "      & [  831.1, 627.7 ] & 129276.20 &  68.01028 &  8.959 &  1.220 &  2.514 &  0.362 & \multicolumn{2}{c}{} & \multicolumn{2}{c}{}  & \\
028-34-Q &      "      & [  849.9, 576.5 ] & 129239.40 &  67.94765 &  7.351 &  1.060 &  2.449 &  0.378 & \multicolumn{2}{c}{} & \multicolumn{2}{c}{}  & 028-35-A\\
028-34-R &      "      & [  867.0, 584.8 ] & 129213.47 &  67.94932 &  5.190 &  0.841 &  2.224 &  0.426 & \multicolumn{2}{c}{} & \multicolumn{2}{c}{}  & 028-35-C\\
028-34-S &      "      & [  870.1, 399.2 ] & 129184.01 &  67.74872 &  8.040 &  1.192 &  4.101 &  0.606 & \multicolumn{2}{c}{} & \multicolumn{2}{c}{}  & 028-35-D\\
028-34-T &      "      & [  876.2, 869.4 ] & 129239.42 &  68.24983 &  7.877 &  1.065 &  2.716 &  0.376 & \multicolumn{2}{c}{} & \multicolumn{2}{c}{}  & \\
028-34-U &      "      & [  876.0, 443.1 ] & 129180.23 &  67.79342 &  6.512 &  0.926 &  2.977 &  0.444 & \multicolumn{2}{c}{} & \multicolumn{2}{c}{}  & 028-35-E\\
028-34-V &      "      & [  879.9, 444.1 ] & 129174.25 &  67.79279 &  5.032 &  0.671 &  5.859 &  0.874 & \multicolumn{2}{c}{} & \multicolumn{2}{c}{}  & 028-35-G\\
028-34-W &      "      & [  931.9, 212.9 ] & 129062.79 &  67.52141 & 12.410 &  1.824 &  4.078 &  0.706 & \multicolumn{2}{c}{} & \multicolumn{2}{c}{}  & 028-35-K\\
028-34-X &      "      & [  955.5, 382.4 ] & 129046.40 &  67.69430 &  6.108 &  0.911 &  2.342 &  0.386 & \multicolumn{2}{c}{} & \multicolumn{2}{c}{}  & 028-35-L\\
028-34-Y &      "      & [  872.0,  96.2 ] & 129144.10 &  67.42087 & 10.073 &  1.267 &  2.677 &  0.524 & \multicolumn{2}{c}{} & \multicolumn{2}{c}{}  & 028-35-B\\
028-34-Z &      "      & [  878.6, 370.2 ] & 129166.77 &  67.71381 &  9.071 &  1.729 &  0.974 &  6.607 & \multicolumn{2}{c}{} & \multicolumn{2}{c}{}  & 028-35-F\\
028-34-a &      "      & [  887.5, 235.3 ] & 129135.88 &  67.56458 &  4.531 &  1.146 &  2.218 &  0.670 & \multicolumn{2}{c}{} & \multicolumn{2}{c}{}  & 028-35-H\\
028-34-b &      "      & [  910.6,  88.9 ] & 129082.08 &  67.39639 &  5.286 &  0.787 &  2.672 &  0.406 & \multicolumn{2}{c}{} & \multicolumn{2}{c}{}  & 028-35-I\\
028-34-c &      "      & [  920.8, 240.5 ] & 129083.82 &  67.55597 &  9.532 &  2.056 &  4.109 &  0.896 & \multicolumn{2}{c}{} & \multicolumn{2}{c}{}  & 028-35-J\\
028-35-A & N1536498453 & [   14.5, 833.1 ] & 129240.37 &  68.64986 &  6.324 &  1.121 &  2.636 &  0.460 & \multicolumn{2}{c}{} & \multicolumn{2}{c}{}  & 028-34-Q\\
028-35-B &      "      & [   32.2, 344.7 ] & 129144.79 &  68.12313 & 12.461 &  1.518 &  2.659 &  0.359 & \multicolumn{2}{c}{} & \multicolumn{2}{c}{}  & 028-34-Y\\
028-35-C &      "      & [   31.7, 841.6 ] & 129214.46 &  68.65183 &  6.262 &  1.164 &  2.205 &  0.374 & \multicolumn{2}{c}{} & \multicolumn{2}{c}{}  & 028-34-R\\
028-35-D &      "      & [   32.3, 652.0 ] & 129186.04 &  68.45050 & 10.196 &  1.415 &  3.869 &  0.538 & \multicolumn{2}{c}{} & \multicolumn{2}{c}{}  & 028-34-S\\
028-35-E &      "      & [   39.2, 698.3 ] & 129181.71 &  68.49684 &  5.934 &  0.939 &  2.159 &  0.314 & \multicolumn{2}{c}{} & \multicolumn{2}{c}{}  & 028-34-U\\
028-35-F &      "      & [   41.4, 624.1 ] & 129167.76 &  68.41715 & 12.066 &  2.128 &  1.705 &  0.245 & \multicolumn{2}{c}{} & \multicolumn{2}{c}{}  & 028-34-Z\\
028-35-G &      "      & [   43.1, 698.6 ] & 129175.61 &  68.49558 & 10.783 &  1.257 &  5.415 &  0.685 & \multicolumn{2}{c}{} & \multicolumn{2}{c}{}  & 028-34-V\\
028-35-H &      "      & [   48.7, 487.4 ] & 129137.53 &  68.26853 &  4.786 &  1.109 &  2.316 &  0.553 & \multicolumn{2}{c}{} & \multicolumn{2}{c}{}  & 028-34-a\\
028-35-I &      "      & [   70.7, 338.3 ] & 129083.11 &  68.10002 &  8.534 &  1.355 &  2.129 &  0.421 & \multicolumn{2}{c}{} & \multicolumn{2}{c}{}  & 028-34-b\\
028-35-J &      "      & [   82.5, 491.3 ] & 129084.56 &  68.25864 & 17.268 &  2.351 &  2.877 &  0.395 & \multicolumn{2}{c}{} & \multicolumn{2}{c}{}  & 028-34-c\\
028-35-K &      "      & [   93.3, 463.9 ] & 129063.96 &  68.22492 & 10.038 &  1.946 &  1.178 &  3.697 & \multicolumn{2}{c}{} & \multicolumn{2}{c}{}  & 028-34-W\\
028-35-L &      "      & [  118.6, 636.8 ] & 129047.59 &  68.39874 &  8.634 &  1.201 &  3.061 &  0.425 & \multicolumn{2}{c}{} & \multicolumn{2}{c}{}  & 028-34-X\\
028-35-M &      "      & [  180.7,  12.6 ] & 128869.93 &  67.70341 &  6.970 &  0.762 &  2.506 &  0.290 & \multicolumn{2}{c}{} & \multicolumn{2}{c}{}  & \\
028-35-N &      "      & [  187.0,  10.5 ] & 128859.72 &  67.69847 &  8.552 &  1.289 &  2.140 &  0.382 & \multicolumn{2}{c}{} & \multicolumn{2}{c}{}  & \\
028-35-O &      "      & [  190.8, 283.5 ] & 128886.50 &  67.99064 &  8.022 &  1.068 &  2.566 &  0.356 & \multicolumn{2}{c}{} & \multicolumn{2}{c}{}  & \\
028-35-P &      "      & [  239.0, 289.0 ] & 128810.96 &  67.97613 & 10.027 &  1.162 &  3.438 &  0.399 & \multicolumn{2}{c}{} & \multicolumn{2}{c}{}  & \\
028-35-Q &      "      & [  241.9, 682.5 ] & 128859.00 &  68.39634 & 12.064 &  1.957 &  1.512 &  0.946 & \multicolumn{2}{c}{} & \multicolumn{2}{c}{}  & \\
028-35-R &      "      & [  272.8, 728.5 ] & 128816.63 &  68.43250 & 11.375 &  0.936 &  3.003 &  0.248 & \multicolumn{2}{c}{} & \multicolumn{2}{c}{}  & \\
028-35-S &      "      & [  290.0, 727.5 ] & 128789.28 &  68.42434 & 11.223 &  1.152 &  3.237 &  0.331 & \multicolumn{2}{c}{} & \multicolumn{2}{c}{}  & \\
028-35-T &      "      & [  291.5, 835.9 ] & 128802.73 &  68.53923 &  7.294 &  0.902 &  1.942 &  0.355 & \multicolumn{2}{c}{} & \multicolumn{2}{c}{}  & \\
028-35-U &      "      & [  297.5, 589.6 ] & 128757.97 &  68.27390 &  6.899 &  0.517 &  3.159 &  0.237 & \multicolumn{2}{c}{} & \multicolumn{2}{c}{}  & \\
028-35-V &      "      & [  298.9, 663.8 ] & 128766.18 &  68.35267 &  8.817 &  0.992 &  2.688 &  0.310 & \multicolumn{2}{c}{} & \multicolumn{2}{c}{}  & \\
028-35-W &      "      & [  338.1, 326.2 ] & 128658.92 &  67.97410 & 12.382 &  1.452 &  2.053 &  0.320 & \multicolumn{2}{c}{} & \multicolumn{2}{c}{}  & \\
028-35-X &      "      & [  385.8, 962.8 ] & 128672.52 &  68.63555 &  9.494 &  1.036 &  2.658 &  0.298 & \multicolumn{2}{c}{} & \multicolumn{2}{c}{}  & \\
028-35-Y &      "      & [  394.2, 887.7 ] & 128647.76 &  68.55204 & 12.969 &  1.551 &  3.423 &  0.409 & \multicolumn{2}{c}{} & \multicolumn{2}{c}{}  & \\
028-35-Z &      "      & [  829.4, 325.4 ] & 127879.92 &  67.76269 &  8.929 &  1.549 &  2.301 &  0.446 & \multicolumn{2}{c}{} & \multicolumn{2}{c}{}  & \\
028-35-a &      "      & [  885.5, 591.1 ] & 127825.54 &  68.02709 & 16.479 &  1.999 &  2.424 &  0.317 & \multicolumn{2}{c}{} & \multicolumn{2}{c}{}  & 028-36-B\\
028-35-b &      "      & [  887.8, 655.7 ] & 127830.78 &  68.09606 &  8.605 &  0.770 &  1.077 &  4.521 & \multicolumn{2}{c}{} & \multicolumn{2}{c}{}  & 028-36-A\\
028-35-c &      "      & [  894.8, 925.6 ] & 127858.84 &  68.38448 &  5.081 &  0.833 &  1.060 &  7.562 & \multicolumn{2}{c}{} & \multicolumn{2}{c}{}  & \\
028-35-d &      "      & [  903.1, 830.3 ] & 127831.44 &  68.27824 &  7.512 &  1.073 &  2.361 &  0.372 & \multicolumn{2}{c}{} & \multicolumn{2}{c}{}  & \\
028-35-e &      "      & [  915.0, 960.4 ] & 127832.01 &  68.41344 &  6.501 &  1.249 &  2.419 &  0.498 & \multicolumn{2}{c}{} & \multicolumn{2}{c}{}  & \\
028-36-A & N1536498545 & [   44.0, 910.0 ] & 127831.84 &  68.80975 &  8.277 &  1.790 &  2.189 &  0.440 & \multicolumn{2}{c}{} & \multicolumn{2}{c}{}  & 028-35-b\\
028-36-B &      "      & [   41.4, 843.0 ] & 127825.82 &  68.73916 &  7.696 &  0.846 &  2.129 &  0.296 & 12.911 &  0.839 &  1.236 &  0.244 & 028-35-a\\
\hline
031-47-A & N1540681073 & [  272.9, 639.8 ] & 131569.72 & 189.11437 &  8.691 &  0.432 &  2.300 &  0.097 & 10.834 &  0.324 &  0.660 &  0.107 & \\
031-47-B &      "      & [  278.2, 469.7 ] & 131577.29 & 189.20502 &  3.779 &  1.364 &  1.513 &  0.595 &  6.194 &  1.342 &  1.249 &  0.721 & \\
031-47-C &      "      & [  558.9, 402.3 ] & 131959.70 & 189.24041 &  8.435 &  1.087 &  2.300 &  0.319 & \multicolumn{2}{c}{} & \multicolumn{2}{c}{}  & \\
\hline
032-44-A & N1541716008 & [  943.9, 667.0 ] & 128952.82 & 185.26602 &  7.541 &  0.614 &  1.872 &  0.229 & \multicolumn{2}{c}{} & \multicolumn{2}{c}{}  & \\
032-45-A & N1541716180 & [  554.9, 895.7 ] & 129535.28 & 185.65615 &  5.999 &  0.328 &  1.843 &  0.097 &  9.629 &  0.318 &  1.000 &  0.118 & \\
\hline
046-07-A & N1560310219 & [  364.6,  15.4 ] & 131738.53 & 105.14608 &  6.686 &  0.726 &  1.408 &  0.144 & \multicolumn{2}{c}{} & \multicolumn{2}{c}{}  & \\
046-07-B &      "      & [  595.8, 135.4 ] & 131567.15 & 105.14851 &  3.059 &  0.299 &  1.220 &  0.109 &  4.038 &  0.245 &  0.585 &  0.117 & \\
046-07-C &      "      & [  657.9, 129.4 ] & 131521.54 & 105.13701 &  5.277 &  0.573 &  1.224 &  0.136 & \multicolumn{2}{c}{} & \multicolumn{2}{c}{}  & \\
046-07-D &      "      & [  660.9, 597.0 ] & 131513.85 & 105.28514 &  3.715 &  0.346 &  1.743 &  0.147 &  4.687 &  0.274 &  0.921 &  0.156 & \\
046-07-E &      "      & [  667.6, 928.5 ] & 131505.63 & 105.38916 &  3.616 &  0.279 &  1.119 &  0.091 &  5.449 &  0.264 &  0.995 &  0.087 & \\
046-07-F &      "      & [  748.2, 793.4 ] & 131447.60 & 105.33413 &  4.363 &  0.197 &  1.310 &  0.062 &  6.755 &  0.187 &  0.840 &  0.062 & \\
046-07-G &      "      & [  768.9, 653.4 ] & 131433.76 & 105.28656 &  6.405 &  0.633 &  1.398 &  0.138 & \multicolumn{2}{c}{} & \multicolumn{2}{c}{}  & \\
046-07-H &      "      & [  820.7, 476.4 ] & 131397.54 & 105.22233 &  3.868 &  0.292 &  1.054 &  0.071 &  5.291 &  0.249 &  0.541 &  0.081 & \\
046-07-I &      "      & [  828.3, 168.5 ] & 131395.55 & 105.12306 &  4.577 &  0.353 &  1.197 &  0.070 &  4.412 &  0.248 &  0.686 &  0.081 & \\
046-07-J &      "      & [  905.8, 513.7 ] & 131334.48 & 105.22115 &  5.181 &  0.599 &  1.103 &  0.083 &  5.397 &  0.386 &  0.337 &  0.109 & \\
046-09-A & N1560310460 & [  338.0, 190.2 ] & 130169.59 & 105.93428 &  3.799 &  0.597 &  1.665 &  0.191 &  3.929 &  0.403 &  0.782 &  0.230 & \\
046-09-B &      "      & [  600.5, 717.2 ] & 129963.79 & 106.06617 &  5.881 &  0.698 &  1.138 &  0.129 & \multicolumn{2}{c}{} & \multicolumn{2}{c}{}  & \\
046-09-C &      "      & [  934.5, 890.6 ] & 129705.45 & 106.06886 &  3.182 &  0.321 &  1.267 &  0.145 &  4.988 &  0.296 &  0.742 &  0.128 & \\
046-10-A & N1560310609 & [  238.2, 236.3 ] & 129435.91 & 106.42638 &  3.413 &  0.324 &  1.086 &  0.095 &  4.269 &  0.268 &  0.824 &  0.100 & \\
046-10-B &      "      & [  260.0, 628.6 ] & 129416.42 & 106.55626 &  3.200 &  0.328 &  1.233 &  0.130 &  4.864 &  0.305 &  0.046 &  0.135 & \\
046-10-C &      "      & [  268.4, 160.9 ] & 129412.57 & 106.39542 &  3.440 &  0.526 &  1.532 &  0.184 &  3.771 &  0.369 &  0.757 &  0.211 & \\
046-10-D &      "      & [  346.7, 607.1 ] & 129348.29 & 106.53413 &  6.721 &  0.694 &  1.391 &  0.144 & \multicolumn{2}{c}{} & \multicolumn{2}{c}{}  & \\
046-10-E &      "      & [  529.5, 351.1 ] & 129205.44 & 106.41515 &  9.647 &  0.863 &  1.206 &  0.112 & \multicolumn{2}{c}{} & \multicolumn{2}{c}{}  & \\
046-10-F &      "      & [  562.3, 772.7 ] & 129177.39 & 106.55358 &  3.960 &  0.478 &  1.576 &  0.137 &  4.654 &  0.332 & -0.148 &  0.173 & \\
046-10-G &      "      & [  580.1, 943.0 ] & 129162.79 & 106.60860 &  8.746 &  0.803 &  1.373 &  0.126 & \multicolumn{2}{c}{} & \multicolumn{2}{c}{}  & \\
046-10-H &      "      & [  596.3, 699.6 ] & 129150.87 & 106.52278 &  6.418 &  0.593 &  1.369 &  0.128 & \multicolumn{2}{c}{} & \multicolumn{2}{c}{}  & \\
046-10-I &      "      & [  605.7, 913.7 ] & 129142.64 & 106.59423 &  3.977 &  0.436 &  1.495 &  0.129 &  4.964 &  0.325 &  0.063 &  0.156 & \\
046-10-J &      "      & [  619.5, 584.2 ] & 129133.07 & 106.47936 &  5.548 &  0.458 &  1.556 &  0.129 & \multicolumn{2}{c}{} & \multicolumn{2}{c}{}  & \\
046-10-K &      "      & [  632.9, 147.3 ] & 129124.96 & 106.32726 &  5.603 &  0.490 &  1.373 &  0.116 &  8.339 &  0.420 &  0.563 &  0.109 & \\
046-10-L &      "      & [  647.7, 928.8 ] & 129109.45 & 106.59219 &  4.554 &  0.445 &  1.145 &  0.098 &  6.010 &  0.363 &  0.490 &  0.112 & \\
046-10-M &      "      & [  655.0, 599.6 ] & 129104.93 & 106.47849 &  7.253 &  0.480 &  1.693 &  0.112 & \multicolumn{2}{c}{} & \multicolumn{2}{c}{}  & \\
046-10-N &      "      & [  658.2, 316.2 ] & 129103.84 & 106.38082 &  4.270 &  0.520 &  1.219 &  0.094 &  4.312 &  0.333 &  0.401 &  0.126 & \\
046-10-O &      "      & [  736.8, 371.7 ] & 129041.31 & 106.38614 &  6.771 &  0.769 &  1.032 &  0.151 & \multicolumn{2}{c}{} & \multicolumn{2}{c}{}  & \\
046-10-P &      "      & [  849.8, 907.2 ] & 128949.60 & 106.55011 &  4.519 &  0.578 &  1.199 &  0.150 &  6.250 &  0.494 &  0.771 &  0.150 & \\
046-10-Q &      "      & [  926.8, 424.8 ] & 128890.52 & 106.37124 &  4.175 &  0.364 &  1.435 &  0.108 &  4.641 &  0.269 &  0.893 &  0.113 & \\
046-10-R &      "      & [  948.0, 486.2 ] & 128873.38 & 106.38866 &  3.065 &  0.258 &  1.027 &  0.094 &  5.587 &  0.269 &  0.424 &  0.098 & \\
046-10-S &      "      & [  959.2, 538.8 ] & 128864.30 & 106.40482 &  5.525 &  0.505 &  1.271 &  0.086 &  5.393 &  0.348 &  0.690 &  0.102 & \\
046-10-T &      "      & [  968.4, 237.1 ] & 128858.60 & 106.29934 &  4.881 &  0.301 &  0.921 &  0.053 &  6.844 &  0.274 &  0.850 &  0.062 & \\
046-10-U &      "      & [  976.4,  96.0 ] & 128853.19 & 106.24928 &  3.850 &  0.316 &  1.122 &  0.082 &  5.219 &  0.265 &  0.476 &  0.094 & \\
046-11-A & N1560310728 & [   41.9, 635.3 ] & 128757.17 & 106.92954 &  4.790 &  0.307 &  1.659 &  0.078 &  5.413 &  0.210 &  0.512 &  0.095 & \\
046-11-B &      "      & [  203.6, 442.2 ] & 128627.78 & 106.83356 &  4.162 &  0.294 &  1.172 &  0.077 &  5.768 &  0.251 &  0.379 &  0.085 & \\
046-11-C &      "      & [  317.6, 472.2 ] & 128535.77 & 106.82362 &  4.970 &  0.274 &  1.488 &  0.074 &  5.737 &  0.208 &  0.963 &  0.075 & \\
046-11-D &      "      & [  521.0, 687.2 ] & 128370.92 & 106.86240 &  6.210 &  0.552 &  1.145 &  0.111 & \multicolumn{2}{c}{} & \multicolumn{2}{c}{}  & \\
046-12-A & N1560310846 & [  426.4, 234.2 ] & 127593.48 & 107.03287 &  5.086 &  0.577 &  1.774 &  0.201 & \multicolumn{2}{c}{} & \multicolumn{2}{c}{}  & \\
046-12-B &      "      & [  444.3, 636.9 ] & 127578.09 & 107.17411 &  7.115 &  0.667 &  2.052 &  0.192 & \multicolumn{2}{c}{} & \multicolumn{2}{c}{}  & \\
046-12-C &      "      & [  458.2, 228.3 ] & 127567.19 & 107.02475 &  6.654 &  0.769 &  1.350 &  0.152 & \multicolumn{2}{c}{} & \multicolumn{2}{c}{}  & \\
046-12-D &      "      & [  514.2, 613.6 ] & 127520.28 & 107.15270 & 10.116 &  0.499 &  2.995 &  0.144 & \multicolumn{2}{c}{} & \multicolumn{2}{c}{}  & \\
046-12-E &      "      & [  709.9, 591.3 ] & 127358.20 & 107.10804 &  5.527 &  0.677 &  1.417 &  0.120 &  5.776 &  0.444 &  0.533 &  0.151 & \\
046-12-F &      "      & [  778.9, 336.2 ] & 127301.29 & 107.00302 &  2.929 &  0.371 &  1.458 &  0.146 &  3.537 &  0.271 &  0.451 &  0.173 & \\
\end{longtable}
\end{scriptsize}
\end{landscape}

\typeout{get arXiv to do 4 passes: Label(s) may have changed. Rerun}

\begin{thebibliography}{}

\bibitem[{Colwell} et~al., 2006]{Colwell06}
{Colwell}, J.~E., {Esposito}, L.~W., and {Srem{\v c}evi{\'c}}, M. (2006).
\newblock {Self-gravity wakes in Saturn's A ring measured by stellar
  occultations from Cassini}.
\newblock {\em \grl}, 33:L07201.

\bibitem[{Colwell} et~al., 2007]{Colwell07}
{Colwell}, J.~E., {Esposito}, L.~W., {Srem{\v c}evi{\'c}}, M., {Stewart},
  G.~R., and {McClintock}, W.~E. (2007).
\newblock {Self-gravity wakes and radial structure of Saturn's B ring}.
\newblock {\em Icarus}, 190:127--144.

\bibitem[{Dones} et~al., 1993]{Dones93}
{Dones}, L., {Cuzzi}, J.~N., and {Showalter}, M.~R. (1993).
\newblock {Voyager photometry of Saturn's A Ring}.
\newblock {\em Icarus}, 105:184--215.

\bibitem[{Esposito} et~al., 2007]{Espo07}
{Esposito}, L.~W., {Mienke}, B.~K., {Colwell}, J.~E., {Nicholson}, P.~D., and
  {Hedman}, M.~M. (2007).
\newblock {Moonlets and clumps in Saturn's F ring}.
\newblock ~\textit{Icarus}, in press.

\bibitem[{French} et~al., 2007]{French07}
{French}, R.~G., {Salo}, H., {McGhee}, C.~A., and {Dones}, L. (2007).
\newblock {HST observations of azimuthal asymmetry in Saturn's rings}.
\newblock {\em Icarus}, 189:493--522.

\bibitem[{Hedman} et~al., 2007]{Hedman07}
{Hedman}, M.~M., {Nicholson}, P.~D., {Salo}, H., {Wallis}, B.~D., {Buratti},
  B.~J., {Baines}, K.~H., {Brown}, R.~H., and {Clark}, R.~N. (2007).
\newblock {Self-gravity wake structures in Saturn's A ring revealed by Cassini
  VIMS}.
\newblock {\em \aj}, 133:2624--2629.

\bibitem[{Hedman} et~al., 2005]{Hedman05}
{Hedman}, M.~M., {Nicholson}, P.~D., and {Wallis}, B.~D. (2005).
\newblock {Cassini-VIMS Observations of Stellar Occultations by Saturn's
  Rings}.
\newblock {\em AGU Fall Meeting Abstracts}, pages P31D--02.

\bibitem[{Jacobson} et~al., 2006]{Jake06}
{Jacobson}, R.~A., {Antreasian}, P.~G., {Bordi}, J.~J., {Criddle}, K.~E.,
  {Ionasescu}, R., {Jones}, J.~B., {Mackenzie}, R.~A., {Meek}, M.~C.,
  {Parcher}, D., {Pelletier}, F.~J., {Owen}, W.~M., {Roth}, D.~C., {Roundhill},
  I.~M., and {Stauch}, J.~R. (2006).
\newblock {The gravity field of the Saturnian system from satellite
  observations and spacecraft tracking data}.
\newblock {\em \aj}, 132:2520--2526.

\bibitem[{Julian} and {Toomre}, 1966]{JT66}
{Julian}, W.~H. and {Toomre}, A. (1966).
\newblock {Non-axisymmetric responses of differentially rotating disks of
  stars}.
\newblock {\em \apj}, 146:810--830.

\bibitem[{Lewis} and {Stewart}, 2006]{LewisDPS06}
{Lewis}, M.~C. and {Stewart}, G.~R. (2006).
\newblock {Simulating the Keeler Gap in Saturn's rings: Wake and edge
  dynamics}.
\newblock {\em AAS/Division for Planetary Sciences Meeting Abstracts},
  38:42.05.

\bibitem[{Lewis} and {Stewart}, 2007a]{LS07}
{Lewis}, M.~C. and {Stewart}, G.~R. (2007a).
\newblock {Features around embedded moonlets in Saturn's rings: The role of
  self-gravity and particle size distributions}.
\newblock ~\textit{Icarus}, submitted.

\bibitem[{Lewis} and {Stewart}, 2007b]{LewisDDA07}
{Lewis}, M.~C. and {Stewart}, G.~R. (2007b).
\newblock {Impact of size distributions and self-gravity on structures around
  moonlets In rings}.
\newblock {\em AAS/Division of Dynamical Astronomy Meeting Abstracts},
  38:12.01.

\bibitem[{Murray} and {Dermott}, 1999]{MD99}
{Murray}, C.~D. and {Dermott}, S.~F. (1999).
\newblock {\em {Solar System Dynamics}}.
\newblock Cambridge Univ. Press, Cambridge.

\bibitem[{Nicholson} et~al., 2007]{NichSOI07}
{Nicholson}, P.~D., {Hedman}, M.~M., {Clark}, R.~N., {Showalter}, M.~R.,
  {Cruikshank}, D.~P., {Cuzzi}, J.~N., {Filacchione}, G., {Capaccioni}, F.,
  {Cerroni}, P., {Hansen}, G.~B., {Sicardy}, B., {Drossart}, P., {Brown},
  R.~H., {Buratti}, B.~J., {Baines}, K.~H., and {Coradini}, A. (2007).
\newblock {A close look at Saturn's rings with Cassini VIMS}.
\newblock ~\textit{Icarus}, in press.

\bibitem[{Porco} et~al., 2005]{Porco05}
{Porco}, C.~C., {Baker}, E., {Barbara}, J., {Beurle}, K., {Brahic}, A.,
  {Burns}, J.~A., {Charnoz}, S., {Cooper}, N., {Dawson}, D.~D., {Del Genio},
  A.~D., {Denk}, T., {Dones}, L., {Dyudina}, U., {Evans}, M.~W., {Giese}, B.,
  {Grazier}, K., {Helfenstein}, P., {Ingersoll}, A.~P., {Jacobson}, R.~A.,
  {Johnson}, T.~V., {McEwen}, A., {Murray}, C.~D., {Neukum}, G., {Owen}, W.~M.,
  {Perry}, J., {Roatsch}, T., {Spitale}, J., {Squyres}, S., {Thomas}, P.,
  {Tiscareno}, M., {Turtle}, E., {Vasavada}, A.~R., {Veverka}, J., {Wagner},
  R., and {West}, R. (2005).
\newblock {Cassini imaging science: Initial results on Saturn's rings and small
  satellites}.
\newblock {\em Science}, 307:1226--1236.

\bibitem[{Porco} et~al., 2007a]{PorcoSci07}
{Porco}, C.~C., {Thomas}, P.~C., {Weiss}, J.~W., and {Richardson}, D.~C.
  (2007a).
\newblock {Saturn's small satellites: Clues to their origins}.
\newblock {\em Science}, 318:1602--1607.

\bibitem[{Porco} et~al., 2007b]{PorcoPhot07}
{Porco}, C.~C., {Weiss}, J.~W., {Richardson}, D.~C., {Dones}, L., {Quinn}, T.,
  and {Throop}, H. (2007b).
\newblock {Simulations of the dynamical and light-scattering behavior of
  Saturn's rings and the derivation of ring particle and disk properties}.
\newblock ~\textit{Astron. J.}, submitted.

\bibitem[{Porco} et~al., 2004]{PorcoSSR04}
{Porco}, C.~C., {West}, R.~A., {Squyres}, S., {McEwen}, A., {Thomas}, P.,
  {Murray}, C.~D., {Delgenio}, A., {Ingersoll}, A.~P., {Johnson}, T.~V.,
  {Neukum}, G., {Veverka}, J., {Dones}, L., {Brahic}, A., {Burns}, J.~A.,
  {Haemmerle}, V., {Knowles}, B., {Dawson}, D., {Roatsch}, T., {Beurle}, K.,
  and {Owen}, W. (2004).
\newblock {Cassini imaging science: Instrument characteristics and anticipated
  scientific investigations at Saturn}.
\newblock {\em Space Sci. Rev.}, 115:363--497.

\bibitem[{Press} et~al., 1992]{NumericalRecipes}
{Press}, W.~H., {Teukolsky}, S.~A., {Vetterling}, W.~T., and {Flannery}, B.~P.
  (1992).
\newblock {\em Numerical Recipes in C: The Art of Scientific Computing}.
\newblock Cambridge University Press, Cambridge.
\newblock (http://www.numerical-recipes.com).

\bibitem[{Salo}, 1995]{Salo95}
{Salo}, H. (1995).
\newblock {Simulations of dense planetary rings III. Self-gravitating identical
  particles.}
\newblock {\em Icarus}, 117:287--312.

\bibitem[{Sei{\ss}} et~al., 2005]{Seiss05}
{Sei{\ss}}, M., {Spahn}, F., {Srem{\v c}evi{\'c}}, M., and {Salo}, H. (2005).
\newblock {Structures induced by small moonlets in Saturn's rings: Implications
  for the Cassini Mission}.
\newblock {\em \grl}, 32:L11205.

\bibitem[{Showalter} et~al., 1986]{Show86}
{Showalter}, M.~R., {Cuzzi}, J.~N., {Marouf}, E.~A., and {Esposito}, L.~W.
  (1986).
\newblock {Satellite ``wakes'' and the orbit of the Encke Gap moonlet}.
\newblock {\em Icarus}, 66:297--323.

\bibitem[{Spahn} and {Srem{\v c}evi{\'c}}, 2000]{SS00}
{Spahn}, F. and {Srem{\v c}evi{\'c}}, M. (2000).
\newblock {Density patterns induced by small moonlets in Saturn's rings?}
\newblock {\em \aap}, 358:368--372.

\bibitem[{Srem{\v c}evi{\'c}} et~al., 2007]{Sremcevic07}
{Srem{\v c}evi{\'c}}, M., {Schmidt}, J., {Salo}, H., {Sei{\ss}}, M., {Spahn},
  F., and {Albers}, N. (2007).
\newblock {A belt of moonlets in Saturn's A ring}.
\newblock {\em \nat}, 449:1019--1021.

\bibitem[{Srem{\v c}evi{\'c}} et~al., 2002]{SSD02}
{Srem{\v c}evi{\'c}}, M., {Spahn}, F., and {Duschl}, W.~J. (2002).
\newblock {Density structures in perturbed thin cold discs}.
\newblock {\em \mnras}, 337:1139--1152.

\bibitem[{Tiscareno} et~al., 2006]{Propellers06}
{Tiscareno}, M.~S., {Burns}, J.~A., {Hedman}, M.~M., {Porco}, C.~C., {Weiss},
  J.~W., {Dones}, L., {Richardson}, D.~C., and {Murray}, C.~D. (2006).
\newblock {100-metre-diameter moonlets in Saturn's A Ring from observations of
  ``propeller'' structures}.
\newblock {\em \nat}, 440:648--650.

\bibitem[{Tiscareno} et~al., 2007]{soirings}
{Tiscareno}, M.~S., {Burns}, J.~A., {Nicholson}, P.~D., {Hedman}, M.~M., and
  {Porco}, C.~C. (2007).
\newblock {Cassini imaging of Saturn's rings II. A wavelet technique for
  analysis of density waves and other radial structure in the rings}.
\newblock {\em Icarus}, 189:14--34.

\bibitem[{Zebker} et~al., 1985]{Zebker85}
{Zebker}, H.~A., {Marouf}, E.~A., and {Tyler}, G.~L. (1985).
\newblock {Saturn's rings - Particle size distributions for thin layer model}.
\newblock {\em Icarus}, 64:531--548.

\end{thebibliography}
\end{document}